\documentclass[]{pasj01}

\begin{document}
\Received{}
\Accepted{}

\title{The Subaru FMOS Galaxy Redshift Survey (FastSound). 
II. The Emission Line Catalog and Properties of Emission Line Galaxies} 

\author{Hiroyuki~\textsc{Okada},\altaffilmark{1}
  Tomonori~\textsc{Totani},\altaffilmark{1}
  Motonari~\textsc{Tonegawa},\altaffilmark{1}
  Masayuki~\textsc{Akiyama},\altaffilmark{2}
  Gavin~\textsc{Dalton},\altaffilmark{3,4}
  Karl~\textsc{Glazebrook},\altaffilmark{5}
  Fumihide~\textsc{Iwamuro},\altaffilmark{6}
  Kouji~\textsc{Ohta},\altaffilmark{6}
  Naruhisa~\textsc{Takato},\altaffilmark{7}
  Naoyuki~\textsc{Tamura},\altaffilmark{8}
  Kiyoto~\textsc{Yabe},\altaffilmark{8,9}
  Andrew~\textsc{J.~Bunker},\altaffilmark{8,10}
  Tomotsugu~\textsc{Goto},\altaffilmark{11}
  Chiaki~\textsc{Hikage},\altaffilmark{8,12}
  Takashi~\textsc{Ishikawa},\altaffilmark{6}
  Teppei~\textsc{Okumura},\altaffilmark{8}
  and Ikkoh~\textsc{Shimizu}\altaffilmark{1}
}

\altaffiltext{1}{Department of Astronomy, School of Science, The University of Tokyo, 7-3-1 Hongo, Bunkyo-ku, Tokyo 113-0033, Japan}
\altaffiltext{2}{Astronomfical Institute, Faculty of Science, Tohoku University, 6-3 Aramaki, Aoba-ku, Sendai, Miyagi 980-8578, Japan}
\altaffiltext{3}{Astrophysics, Department of Physics, Denys Wilkinson Building, Keble Road, Oxford OX1 3RH, U.K.}
\altaffiltext{4}{RALSpace, STFC Rutherford Appleton Laboratory, HSIC, Oxford OX11 0QX, UK}
\altaffiltext{5}{Centre for Astrophysics \& Supercomputing, Swinburne University of Technology, P.O. Box 218, Hawthorn, VIC 3122, Australia}
\altaffiltext{6}{Department of Astronomy, Kyoto University, Sakyo-ku, Kyoto 606-8502, Japan}
\altaffiltext{7}{Subaru Telescope, National Astronomical Observatory of Japan, 650 North A`ohoku Pl., Hilo, Hawaii 96720, USA}
\altaffiltext{8}{Kavli Institute for the Physics and Mathematics of the Universe (WPI), Todai Institutes for Advanced Study, the University of Tokyo, 5-1-5 Kashiwanoha, Kashiwa, 277-8583, Japan}
\altaffiltext{9}{National Astronomical Observatory of Japan, Mitaka, Tokyo 181-8588, Japan}
\altaffiltext{10}{Department of Physics, University of Oxford, Keble Road, Oxford, OX13RH, UK}
\altaffiltext{11}{Institute of Astronomy, National Tsing Hua University, No. 101, Section 2, Kuang-Fu Road, Hsinchu, Taiwan 30013}
\altaffiltext{12}{Kobayashi-Maskawa Institute for the Origin of Particles and the Universe (KMI), Nagoya University, 464-8602, Japan}

\email{okada@astron.s.u-tokyo.ac.jp}


\maketitle

\begin{abstract}
We present basic properties of $\sim$3,300 emission line galaxies
detected by the FastSound survey, which are mostly H$\alpha$ emitters
at $z \sim$ 1.2--1.5 in the total area of about 20 deg$^2$, with the
H$\alpha$ flux sensitivity limit of $\sim 1.6 \times 10^{-16}~\rm erg
\ cm^{-2} s^{-1}$ at 4.5 sigma. This paper presents the catalogs of
the FastSound emission lines and galaxies, which is open to the
public.  We also present basic properties of typical FastSound
H$\alpha$ emitters, which have H$\alpha$ luminosities of
$10^{41.8}$--$10^{43.3}$~erg/s, SFRs of 20--500 $M_\odot$/yr, and
stellar masses of $10^{10.0}$--$10^{11.3}$ $M_\odot$. The 3D
distribution maps for the four fields of CFHTLS W1--4 are presented,
clearly showing large scale clustering of galaxies at the scale of
$\sim$ 100--600 comoving Mpc.  Based on 1,105 galaxies with detections
of multiple emission lines, we estimate that contamination of
non-H$\alpha$ lines is about 4\% in the single-line emission galaxies,
which are mostly [OIII]$\lambda$5007.  This contamination fraction is
also confirmed by the stacked spectrum of all the FastSound spectra,
in which H$\alpha$, [NII]$\lambda \lambda$6548,6583, [SII]$\lambda
\lambda$6717,6731, and [OI]$\lambda \lambda$6300,6364 are seen.
\end{abstract}

\section{Introduction}\label{Introduction}

The FastSound survey is a galaxy redshift survey using the
near-infrared (NIR) Fiber Multi-Object Spectrograph (FMOS) mounted on
the Subaru Telescope \citep{Kimura}, which detected $\sim$4,000 line
emitters at 1.43--1.67 $\mu$m with a line sensitivity limit of $\sim
1.6 \times 10^{-16}\ \rm erg \ cm^{-2} s^{-1}$ (\cite{Tonegawac},
hereafter Paper I). The majority of these are expected to be H$\alpha$
emitters at $1.19 < z < 1.54$, because emitters at this flux level
have very high H$\alpha$/[OIII] ratios and therefore very low [OIII]
contamination \citep{Colbert}.  The FastSound observations have
already been completed, with the total surveyed area of 20.6 deg$^2$
in the four fields of the CFHTLS Wide Survey \citep{Goranova}.  The
primary scientific objective is to measure the structure growth rate
[i.e., $f(z)\sigma_8(z)$] using the redshift space distortion (RSD),
motivated by the modified gravity scenario for the origin of the
mysterious acceleration of cosmic expansion (see
\cite{Hamilton,Clifton} for reviews).  RSD measurements have been
obtained mainly by redshift surveys at $z < 1$ \citep{Hawkins, Guzzo,
  Blake, Beutler, Bielby, Reid, Samushia, Torre}. FastSound aims to
detect and measure the growth rate in the unexplored redshift range at
a significance of $\sim 4 \sigma$. The FastSound data will also be
useful for general studies on galaxy formation and evolution, like the
past major galaxy redshift surveys at $z \gtrsim 0.5$ in optical and
NIR bands \citep{Wirth, Fevre, Lilly, Coil, Cooper, Brammer, Yabe,
  Newman, Colbert, Kriek}.

This is the second of the FastSound paper series, in which we describe
the FastSound catalog\footnote{The FastSound catalog is publicly
  available at the project website,
  http://www.kusastro.kyoto-u.ac.jp/Fastsound/.}  and present basic
properties of the detected emission lines and their host galaxies.
Line identification is an important issue in FastSound, because most
($\gtrsim$ 70\%) of the detected galaxies have only one emission line,
due to the narrow wavelength range. We expect that the majority of the
FastSound emission lines are H$\alpha$, because of the strength of
H$\alpha$ and our target selection to maximize the probability of
finding H$\alpha$ emitters by photometric estimates of redshift and
star formation rate (SFR).  Indeed, a previous study estimated the
non-H$\alpha$ contamination rate for FastSound to be $\lesssim$10\%
\citep{Tonegawaa} by using the data of the HiZELS narrow-band survey
in NIR \citep{Geach, Sobral09}. 
\citet{Glazebrook05} estimated this
rate to be $\sim$30\% at a similar flux limit from line luminosity
functions, which is consistent with the estimate by \citet{Tonegawaa}
that takes into account the FastSound target selection preferring
strong H$\alpha$ emitters.  Nevertheless, a more precise estimate of
the contamination rate is important for high precision clustering
analyses to derive cosmological information. Therefore we evaluate the
non-H$\alpha$ contamination rate using the FastSound data, based on
the statistics of galaxies with multiple lines and the stacked
spectrum of all FastSound galaxies.

This paper is organized as follows.  The FastSound catalog is
described in \S~\ref{ELcatalog}, and the properties of emission line
galaxies in the catalog are discussed in \S~\ref{Property}.  We
estimate the non-H$\alpha$ contamination in the catalog in
\S~\ref{section:contamination}, followed by conclusions in
\S~\ref{Conclusion}.  Throughout this paper, the standard $\Lambda$CDM
cosmology with the parameters $(\Omega_{\rm m},\Omega_{\Lambda},h)=(0.3,
0.7, 0.7)$ is assumed, where $h = H_0 / (100~{\rm km\ s}^{-1}{\rm
  Mpc}^{-1})$.

\section{The FastSound Catalogs$^1$}\label{ELcatalog}

The FastSound catalog data set consists of the three tables: (1) {\it
  the FoV-information list} for the record and statistics of each FMOS
field-of-view
(FoV) observed in the FastSound survey, 
(2) {\it the galaxy catalog} that
describes the properties and line-detection statistics for all
galaxies selected as the FastSound targets, and (3) {\it the emission
  line catalog} that includes all the emission line candidates
detected by the software FIELD as described in Paper I.  These three
tables are given separately for the four CFHTLS Wide fields observed
by the FastSound survey, and detailed descriptions are given below.

\subsection{The FoV-Information List}
\label{section:FoV-information}

The FoV-information list is provided so that users can get information
about each of the 121 FMOS FoVs observed in the FastSound survey. The
basic properties of each FoV, such as the observation date, FMOS FoV
ID (e.g., W1\_011 meaning the 11th FMOS FoV in the W1 field), and
central coordinates are given. (See Table
\ref{table:catalog_description_FoV} for the list of entries in this
table.) We also provided statistics for the number of galaxies and
emission lines, such as the number of target galaxies sent to the
fiber allocation software ($N_{\rm tar}$), the number of galaxies
actually observed by FMOS fibers ($N_{\rm obs}$), the number of
galaxies with at least one detected emission line at $S/N \ge 4.5$
($N_{\rm elg}$), and the total number of detected emission line
candidates ($N_{\rm el}$).  The statistics of $N_{\rm elg}$ and
$N_{\rm el}$ are given not only for the normal frames but also the
inverted frames\footnote{A normal frame is produced by subtracting the
  sky frame from the object frame, while an inverted frame is by the
  inverted procedure, i.e., sky minus object, with all other reduction
  processes unchanged. All the lines detected in inverted frames
  must be spurious, and the number of these is a good indicator of
  that in normal frames. See Paper I for details.}.  FMOS has two
spectrographs, IRS1 and IRS2, each of which covers about 200
objects. Therefore these numbers are given separately for IRS1 and
IRS2.

The information about observing conditions is also given, such as the
mean seeing. The mean seeing for each FoV is derived from the 2D
optical image of coordinate calibration stars (CCS, typically 17 CCSs
in a FoV) by fitting it with a gaussian profile.  An important
quantity is $\langle f_{\rm obs}\rangle$ calculated using the flux
calibration stars (FCS, typically eight FCSs in a FoV).  For each FCS
we calculated $f_{\rm obs}$, which is the ratio of the uncalibrated
flux within the fiber aperture ($f_{\rm el, raw}$, see \S
\ref{section:emission-line-catalog} below) to the total stellar flux
in the literature (see \S 5 of Paper I).  Then the average in a FoV is
calculated as $\langle f_{\rm obs}\rangle$. The line fluxes $f_{\rm
  el,raw}$ are {\it not} corrected for the observing conditions, and
hence the factor $\langle f_{\rm obs}\rangle^{-1}$ should be
multiplied to get a flux that is calibrated by FCSs and corrected for
the fiber aperture effect (assuming a point source). For extended
sources, further corrections are necessary.  See \S
\ref{section:emission-line-catalog} for the method of calculating the
emission line fluxes of FastSound galaxies used in this paper.

Emission line detection efficiency for a galaxy observed by FMOS is
not uniform over the observed FastSounds fields, because of variable
weather and hardware conditions. This can be estimated by the ratio
$N_{\rm elg}/N_{\rm obs}$. For a galaxy selected as a FastSound
target, the mean probability of being actually observed is $\sim$80\%
(typically, 360 fibers allocated to 440 targets per FoV, see Fig. 5 of
Paper I) in the fiber allocation procedure.  The line detection
efficiency against the target galaxies, including this effect, can be
estimated by the ratio $N_{\rm elg}/N_{\rm tar}$. This efficiency
would be useful for a clustering analysis of the spatial galaxy
distribution.

\subsection{The Galaxy Catalog}

The galaxy catalog provides various information for individual
galaxies, including not only those actually observed by FMOS
but also all the galaxies meeting the selection criteria of the FastSound
target galaxies, which were sent to the fiber allocation software of FMOS
(see Paper I for details).
See Table \ref{table:catalog_description_gal} for the list of
items in this catalog.  As basic information of a galaxy, the
following quantities are given: the CFHTLS-W ID in the catalog of
\citet{Gwyn}, RA, Dec, and CFHTLS optical magnitudes.  The
near-infrared $JK$ magnitudes of the UKIDSS DXS [DR8,
  \citet{Lawrence}] are also added where available (only in W1 and W4
fields). The Galactic extinction by the extinction map of
\citet{Schlegel} is also given, and these magnitudes have been
corrected for this.  

The physical quantities estimated by the photometric SED fittings
performed for target selection are also provided: photometric redshift
($z_{\rm ph}$), SFR, stellar mass ($M_*$), $E(B-V)$, and H$\alpha$
flux estimate (see Paper I for the fitting method).  The median, the
likelihood peak value, and 68\% confidence level (C.L.) lower and upper bounds derived by
the photometric redshift code {\it LePhare} \citep{Arnouts} are given for $z_{\rm
  ph}$, SFR, and $M_*$ in the table.  When the likelihood function
does not have a simple form, the likelihood peak value may be outside
the range between the 68\% C.L. lower and upper limits. 
The 68\% C.L. upper and lower limits for H$\alpha$
flux estimates are based on those for SFR.

The number of FMOS observations for a galaxy, $n_{\rm obs}$, is zero
if no fiber was allocated to it.  In each FoV, observation was done
with only one fiber allocation pattern, and hence most galaxies
observed by FMOS have $n_{\rm obs} = 1$, though it can be two or
larger for some galaxies in FoV overlapping regions.  We also provide
the number of detected emission lines for each galaxy ($n_{\rm el2}$,
see the next section for details).  Since the typical probability of
detecting emission lines from a target galaxy is about 10\% in the
FastSound survey, this catalog includes about 10 times more galaxies
than those in the emission line catalog, and about 90\% of them have
$n_{\rm el2} = 0$.  The reasons for this relatively low detection
efficiency are discussed in \citet{Tonegawaa} and Paper I.

\subsection{The Emission Line Catalog}
\label{section:emission-line-catalog}

\begin{figure}[htp]
\begin{center}
\begin{tabular}{c}
 \begin{minipage}{0.5\hsize}
 \begin{center}
  \includegraphics[width=0.2\paperwidth]{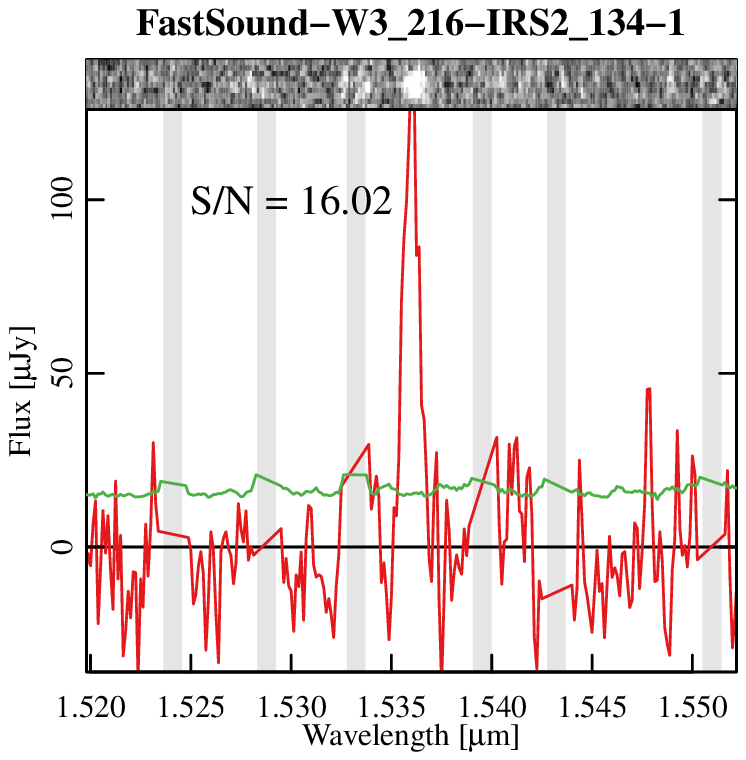}
 \end{center}
 \end{minipage}
 \begin{minipage}{0.5\hsize}
 \begin{center}
  \includegraphics[width=0.2\paperwidth]{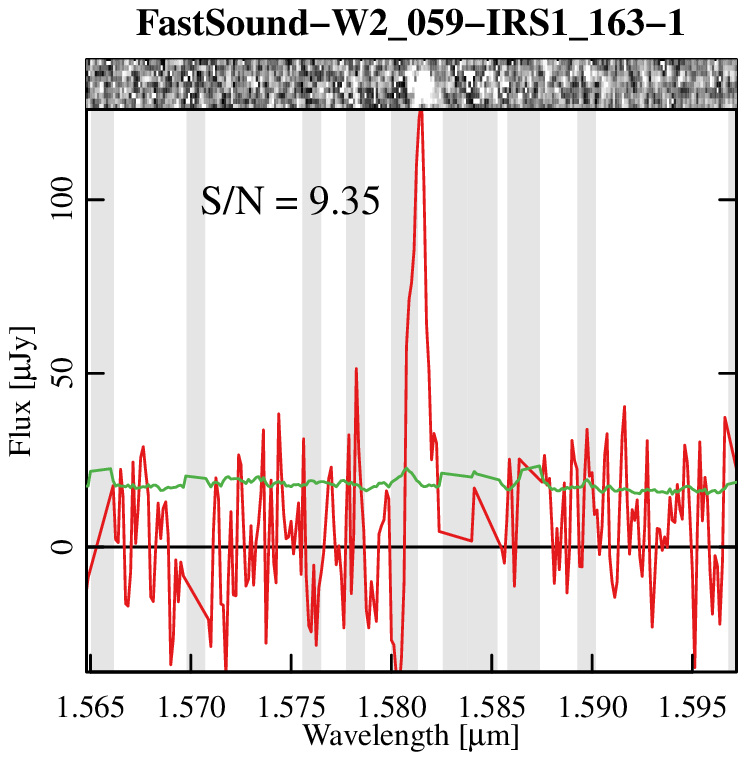}
 \end{center}
 \end{minipage}
\\
 \begin{minipage}{0.5\hsize}
 \begin{center}
  \includegraphics[width=0.2\paperwidth]{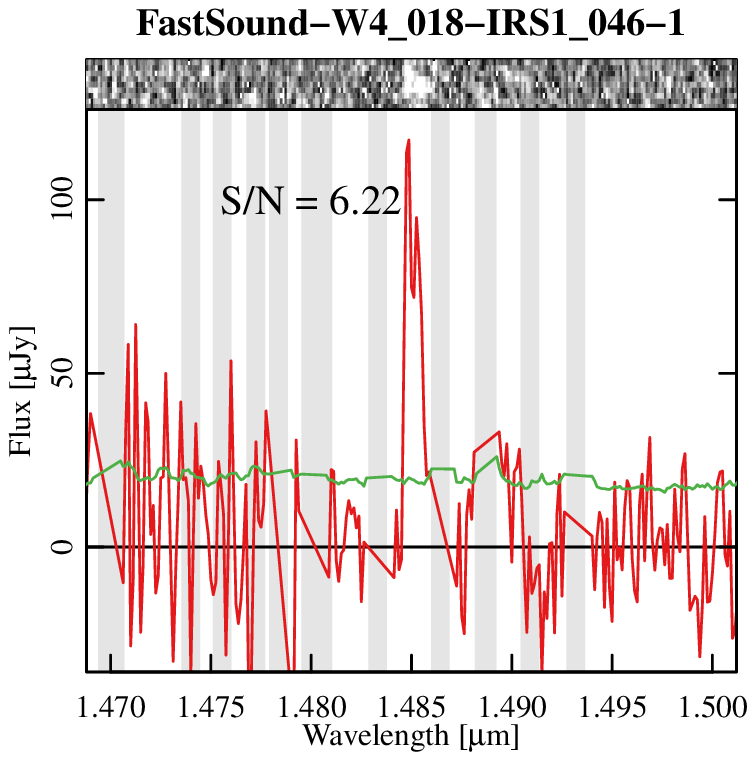}
 \end{center}
 \end{minipage}
 \begin{minipage}{0.5\hsize}
 \begin{center}
  \includegraphics[width=0.2\paperwidth]{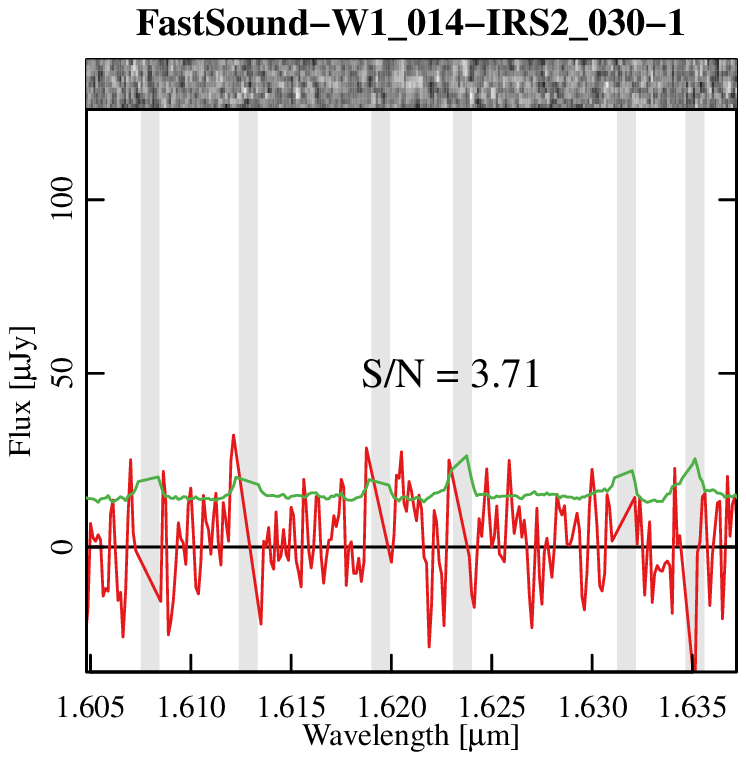}
 \end{center}
 \end{minipage}
\end{tabular}
\end{center}
  \caption{Typical emission line spectra with various $S/N$ values.
    The 2D spectra are shown on the top of each panel.  The grey
    vertical stripes indicate the wavelength ranges corresponding to
    the FMOS OH airglow masks. The green lines show the noise level,
    multiplied by a factor of five for the presentation purpose.  We
    set a threshold of $S/N \geq 4.5$ for the single emission line
    catalog, and lines with $3.0 \leq S/N < 4.5$ are considered only
    when they are accompanied by a line of $S/N \geq 4.5$ in the same
    galaxy with an expected wavelength ratio consistent with known line
    pairs (see \S~\ref{section:multi-line-stat}).  }
\label{fig:individuals}
\end{figure}

The quantities presented in the emission line catalog are summarized
in Table \ref{table:catalog_description_el}.  This catalog includes
all emission line candidates detected by the line detection software
FIELD [\citet{Tonegawab}; Paper I], and the same emission line in a
galaxy may be included as more than one entry in the catalog, if its
host galaxy was observed more than once in the overlapping FastSound
FoVs.  The line candidates are identified by the FastSound ID, which
consists of the FastSound FoV ID, the used spectrograph (IRS1 or 2),
the FMOS fiber ID (1--200 for each of IRS1 and 2), and the line ID
($i$ for the $i$-th line in a spectrum in increasing wavelength
order).  An example of the FastSound ID is
FastSound-W1\_011-IRS1\_009-1.  The line $S/N$ values are also given,
which were calculated by the fixed-width kernel method used in the
line candidate selection (Paper I).  We present typical spectra of
emission line candidates for various $S/N$ values in
Fig. \ref{fig:individuals}.  The emission line catalogs are provided
both for the normal and inverted frames, and the statistics about
spurious noise detection can be examined by using the inverted-frame
catalogs.

As for the physical properties of the lines, the central wavelength
($\lambda$), line flux ($f_{\rm el,raw}$), and line width
($\sigma_\lambda$, not corrected for the instrumental resolution) are
provided, which were calculated by 1D Gaussian fits.  The line flux
was integrated over all wavelengths using the best-fit Gaussian,
including the OH mask regions which represent sharp gaps in the
wavelength coverage. Note that $f_{\rm el,raw}$ in the catalog is not
corrected for the point-source aperture loss due to the fibre (as
determined from the flux calibration stars).

Sometimes wavelengths of line candidates are close to the OH mask
regions.  An OH mask has 7 {\AA} width, and about 300 and 130 masks in
0.8--1.9 $\mu$m were installed for IRS1 and 2, respectively, before
September 2012.  The number for IRS1 is larger because it includes
masks for fainter OH lines. After September 2012, the IRS1 mask plate
has been replaced with a new one that is the same for IRS2, and hence
the OH mask patterns became the same for the two spectrographs (see
S4.3 of Paper I). To give information about nearby OH masking for each
emission line candidate, we provide the OH mask flag parameter (1 if
the line center is within an OH mask width, 0 otherwise), and distance
from the line center to the nearest OH mask borders ($d_{\rm mask}$)
in units of the measured velocity width $\sigma_\lambda$ of the line
candidate. The line centers were determined by a profile fitting, and
sometimes they fall in the mask regions, making the flag 1.  We
provide $d_{\rm mask}$ for both the increasing and decreasing
wavelength directions.  210 line candidates have line centers
overlapped with masked regions (i.e., the flag 1) among the 3,769 line
candidates detected at $S/N \geq 4.5$,

The total (i.e., corrected for the fiber aperture loss) line flux from
a galaxy ($f_{\rm el}$), which is calibrated by the flux standard
stars, is necessary for many studies about galaxy properties.  As
explained in \S \ref{section:FoV-information}, these corrections can
easily be done by using $\langle f_{\rm obs} \rangle$ for point
sources. For extended galaxies, a perfect correction is difficult and
it would be generally model dependent, but here we calculate $f_{\rm
  el}$ as follows. First $f_{\rm el,raw}$ is multiplied by the factor
of $\langle f_{\rm obs} \rangle^{-1}$, to convert it into the total
flux which would be correct if it was a point source. Then we need a
further correction to take into account sizes of extended galaxies.
For this correction we first estimate the calibrated line flux of a
galaxy within the fiber aperture (1.2 arcsec diameter), by multiplying
the fiber covering fraction $c_*$ for a point source, i.e., $c_*
\langle f_{\rm obs} \rangle^{-1} f_{\rm el,raw}$. The factor $c_*$
should be discriminated from $\langle f_{\rm obs}\rangle$; the former
is just the fraction of flux within the fiber aperture for point
sources, while the latter is the combined factor of the fiber aperture
loss and flux reduction by observing conditions such as weather. We
calculated $c_*$ by assuming a two-dimensional Gaussian profile with
the seeing FWHM of each FoV. Then the final step is to correct it by
using the fiber covering fraction $c_{\rm gal}$ for extended galaxies.
Though $c_{\rm gal}$ should depend on the surface brightness profile
of individual galaxies and seeing, we adopt a constant value of
$c_{\rm gal} = 0.47$, as estimated by \citet{Tonegawaa} based on the
image size of typical FastSound galaxies. Here we ignored the effect
of seeing variation in different FastSound FoVs on $c_{\rm gal}$, but
it should be smaller than that for $c_*$ because galaxies are extended
sources. Taking into account the Galactic extinction, the final flux
is then obtained as $f_{\rm el} = 10^{0.4 A_{\rm MW}(\lambda)} c_{\rm
  gal}^{-1} \, c_* \, \langle f_{\rm obs} \rangle^{-1} \, f_{\rm
  el,raw}$, where $A_{\rm MW}(\lambda)$ is the Galactic extinction
magnitude at the observed wavelength, calculated from the Galactic
$E(B-V)_{\rm MW}$ in the catalog and the Milky-Way extinction curve of
\citet{Cardelli}.  The flux $f_{\rm el}$ is also given in the emission
line catalog.

Sometimes lines are blended with nearby lines by line widths
comparable with line separations. In the FastSound line detection
process, line $S/N$ peaks are treated as independent lines when their
separation is more than 20 pixels ($\sim 22$~{\AA}), corresponding to
a velocity difference of $\sim 440$ km/s. Therefore, there is no
blending among different line candidates in the catalog if their
velocity dispersions are typical for normal galaxies ($\lesssim$ 200
km/s).  (Note that some line pairs in the catalog may have line
separations less than 20 pixels in terms of the line central
wavelength $\lambda$, which is estimated by the 1D Gaussian fit and
different from the $S/N$ peak location in the line detection process.)
When line separations are smaller than 20 pixels, such as the
[OII]$\lambda$3727 doublet (separation of about 220 km/s), these lines
are treated as a single line in our sample. As information about
nearby lines, we provided $d_{\rm line}$ in the catalog, which is the
distance in wavelength to the next independent line in the spectrum,
both in increasing and decreasing wavelength directions.

Two numbers are provided for each line candidate in this catalog, to
describe the statistics about the number of lines in a galaxy. 
$n_{\rm el1}$ is the number of detected emission lines in a galaxy,
from an individual observation in which the line candidate was detected.  If
$n_{\rm el1} \ge 2$, this means that the galaxy has multiple lines at
different wavelengths.  $n_{\rm el2}$ is the sum of
$n_{\rm el1}$ over all the observations for a galaxy, and hence
$n_{\rm el2}$ can be larger than $n_{\rm el1}$ if a galaxy was
observed more than once.  For convenience to select the strongest line
in a galaxy, we provided the rank of the line $S/N$ within a galaxy,
$r_{\rm SN1}$ and $r_{\rm SN2}$, among the $n_{\rm el1}$ and $n_{\rm
  el2}$ line candidates, respectively.  For example, if one is
interested only in the strongest line in a galaxy, choosing only lines
having $r_{\rm SN2} = 1$ gives a complete line sample without
duplications.

\begin{table*}[htp]
\footnotesize
\caption{The quantities presented in the FoV-information list.}
\begin{center}
\label{table:catalog_description_FoV}
\begin{tabular}{ll}
\hline \hline
Column name & Description \\
\hline
OBSDATE				&
Date of observation (for 18:00 Hawaiian Standard Time
on the night during which the FoV was observed) \\
FastSound FoV ID	 
  & e.g., CFHTLSW1\_011 for the 11th FoV in the W1 field \\
RAcent					&	RA [deg] (J2000) of the
  FMOS FoV center \\
DECcent					&	DEC [deg] (J2000) of the 
  FMOS FoV center \\
$N_{\rm tar}$			&	
 Number of selected target galaxies sent to fiber allocation \\ 
$N_{\rm obs}$		&
 Number of galaxies actually observed by FMOS (for IRS1 and IRS2) \\
$N_{\rm el}$ & Total number of emission line candidates (for IRS1 and IRS2) \\
$N_{\rm elg}$ &	
 Number of galaxies with at least one detected emission line candidate 
   (for IRS1 and IRS2) \\
mean seeing		&	Mean seeing (FWHM) during exposure [arcsec] \\
$\langle f_{\rm obs} \rangle$		&
 The correction factor for the uncalibrated flux in the emission line
catalog, for IRS1 and IRS2 \\
\hline \hline
\end{tabular}
\end{center}
\end{table*}

\begin{table*}[tp]
\footnotesize
\caption{The quantities presented in the galaxy catalog.}
\begin{center}
\label{table:catalog_description_gal}
\begin{tabular}{lll}
\hline \hline
Column name & Unit & Description \\
\hline
CFHTLSW ID				&			&	CFHTLS-Wide Object ID of \citet{Gwyn} \\
RA						&[deg]		&	J2000\\
DEC						&[deg]		&	J2000\\
MAG\_AUTO			&			&	
The CFHTLS Wide $u'g'r'i'z'$ magnitudes of \citet{Gwyn}. \\
					&			& 
When available, the $JK$ magnitudes from UKIDSS DXS DR8 are also given. 
\\
						&			&
These have been corrected for the Galactic extinction. \\
$E(B-V)_{\rm MW}$ & & Galactic extinction \citep{Schlegel}\\
$z_{\rm ph}$			&			&
Photometric redshift  \\
$M_*$					&[$M_\odot$]    &
Stellar mass estimated by photometric SED fitting \\
SFR						&[$M_\odot/$yr]&
Star formation rate estimated by photometric SED fitting \\
$E(B-V)$					&			&
Extinction in a galaxy estimated by photometric SED fitting \\ 
$f_{\rm estH\alpha}$			&[erg/cm$^2$/s]&	
H$\alpha$ flux estimated from photometric SED fitting \\
$n_{\rm obs}$			&			&
Number of FMOS observations made for a galaxy
(can be $\ge 2$ in overlapping FoV regions)\\
$n_{\rm el2}$			&			&
Total number of emission line candidates
among all the FMOS observations for a galaxy
(see Table \ref{table:catalog_description_el})
\\
\hline \hline
\end{tabular}
\end{center}
\end{table*}

\begin{table*}[tp]
\footnotesize
\caption{The quantities presented in the emission line catalog.}
\begin{center}
\label{table:catalog_description_el}
\begin{tabular}{lll}
\hline \hline
Column name & Unit & Description \\
\hline
FastSound ID				&			&	
ID including the FastSound FoV ID, spectrograph (IRS1 or 2),
fiber ID, and line ID \\
& & \hspace{5mm} (e.g., FastSound-W1\_011-IRS1\_009-1)\\
CFHTLSW ID\footnotemark[$*$] & & 
CFHTLS-Wide Object ID of \citet{Gwyn} for the line host galaxy \\
$S/N$						&			&
The line detection signal-to-noise ratio derived 
by the line detection software FIELD \\
$\lambda$				&[$\mu$m]	&
Central wavelength of the line \\
$f_{\rm el,raw}$			&[erg/cm$^2$/s]&
Flux of the emission line detected (uncalibrated, within fiber aperture) \\
$f_{\rm el}$			&[erg/cm$^2$/s]&
Total calibrated line flux from a galaxy, corrected for aperture loss
and Galactic extinction \\
$\sigma_\lambda$					&[\AA]		&
The 1$\sigma$ line width of the best-fit Gaussian 
(instrumental spectral resolution not deconvolved)\\
OH mask flag & & 1 if the line center is on the OH mask regions, 0 otherwise \\
$d_{\rm mask}$ & $\sigma_\lambda$ 
  & Distances from the line center to the nearest OH mask borders in 
  increasing/decreasing wavelength directions \\ 
$d_{\rm line}$ 		& [$\mu$m] 		&
  Distances from the line center to that of the nearest line in 
  increasing/decreasing wavelength directions \\ 
$n_{\rm el1}$			&			&
Total number of the emission line candidates in the line host
galaxy, \\
& & \hspace{5mm} in the FMOS observation in which the line was detected \\
$n_{\rm el2}$			&			&
The sum of $n_{\rm el1}$ for all the FMOS observations for the line host 
galaxy \\
$r_{\rm SN1}$			&			&
$S/N$ rank of the emission line among the $n_{\rm el1}$
lines (1 for the highest) \\
$r_{\rm SN2}$			&			&
$S/N$ rank of the emission line among the $n_{\rm el2}$
lines (1 for the highest) \\
\hline \hline
\end{tabular}
\end{center}
\footnotemark[$*$]Other quantities about host galaxies in the galaxy catalog
(e.g., RA, DEC) 
are also given in the electronic version of this table,
but they are not shown here. \\
\end{table*}

\begin{figure*}[p]
\begin{center}
  \includegraphics[width=0.87\paperwidth]{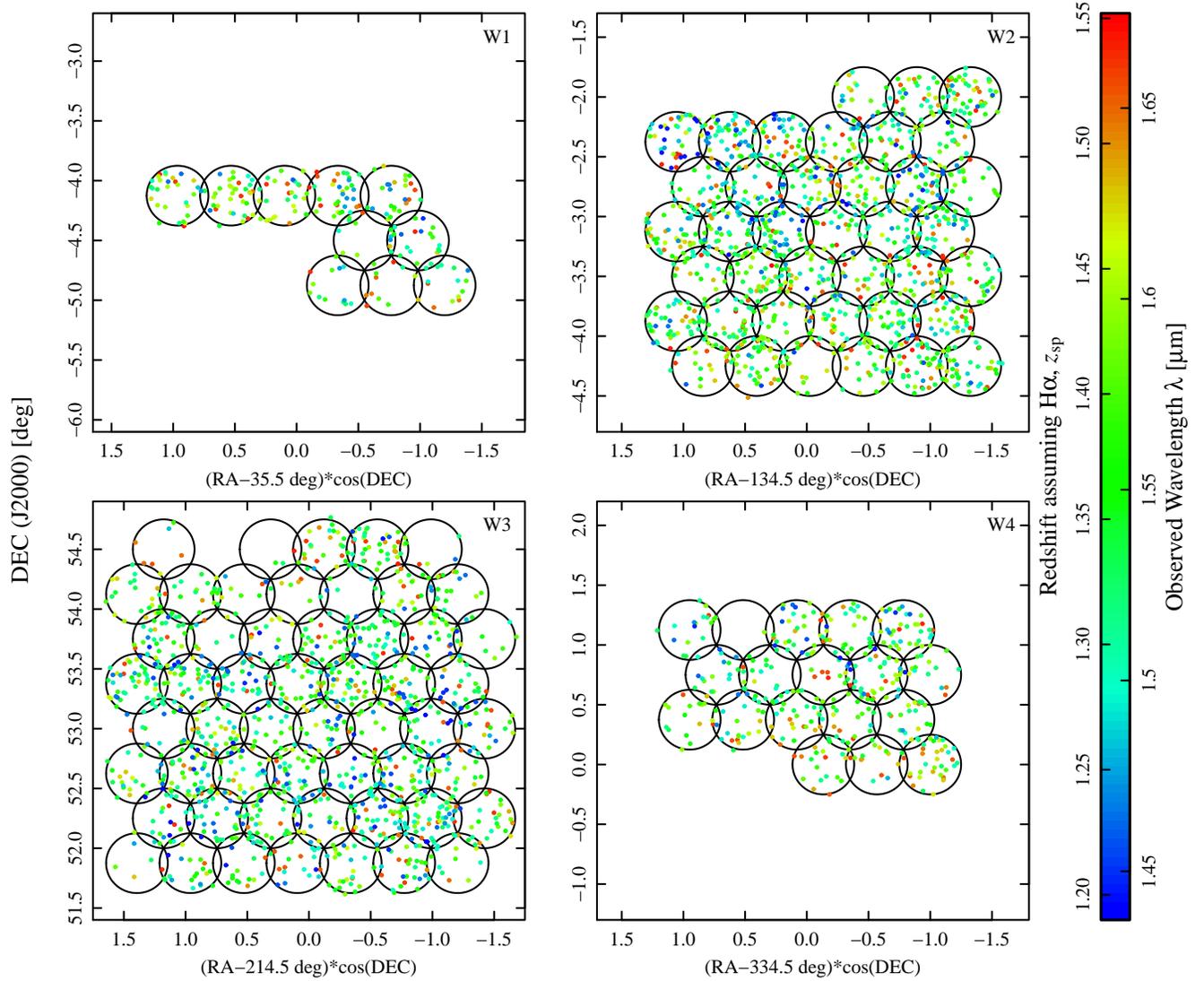}
\end{center}
  \caption{Distribution of galaxies with detected emission lines ($S/N
    > 4.5$) on the celestial sphere in the observed fields CFHTLS
    W1--4.  The color of each dot indicates the observed wavelength of
    the strongest line in a galaxy.  Redshift is also indicated on the
    color scale assuming that the strongest line is H$\alpha$.  Black
    circles represent 30$'$-diameter FMOS FoVs.  (Some galaxies appear
    slightly outside the circles, because FMOS fibers can be allocated
    to objects about 1.7 arcmin outside the 30$'$-diameter FoV.) }
\label{fig:RA_DEC}
\end{figure*}

\begin{figure*}[p]
\begin{center}
  \includegraphics[width=0.87\paperwidth]{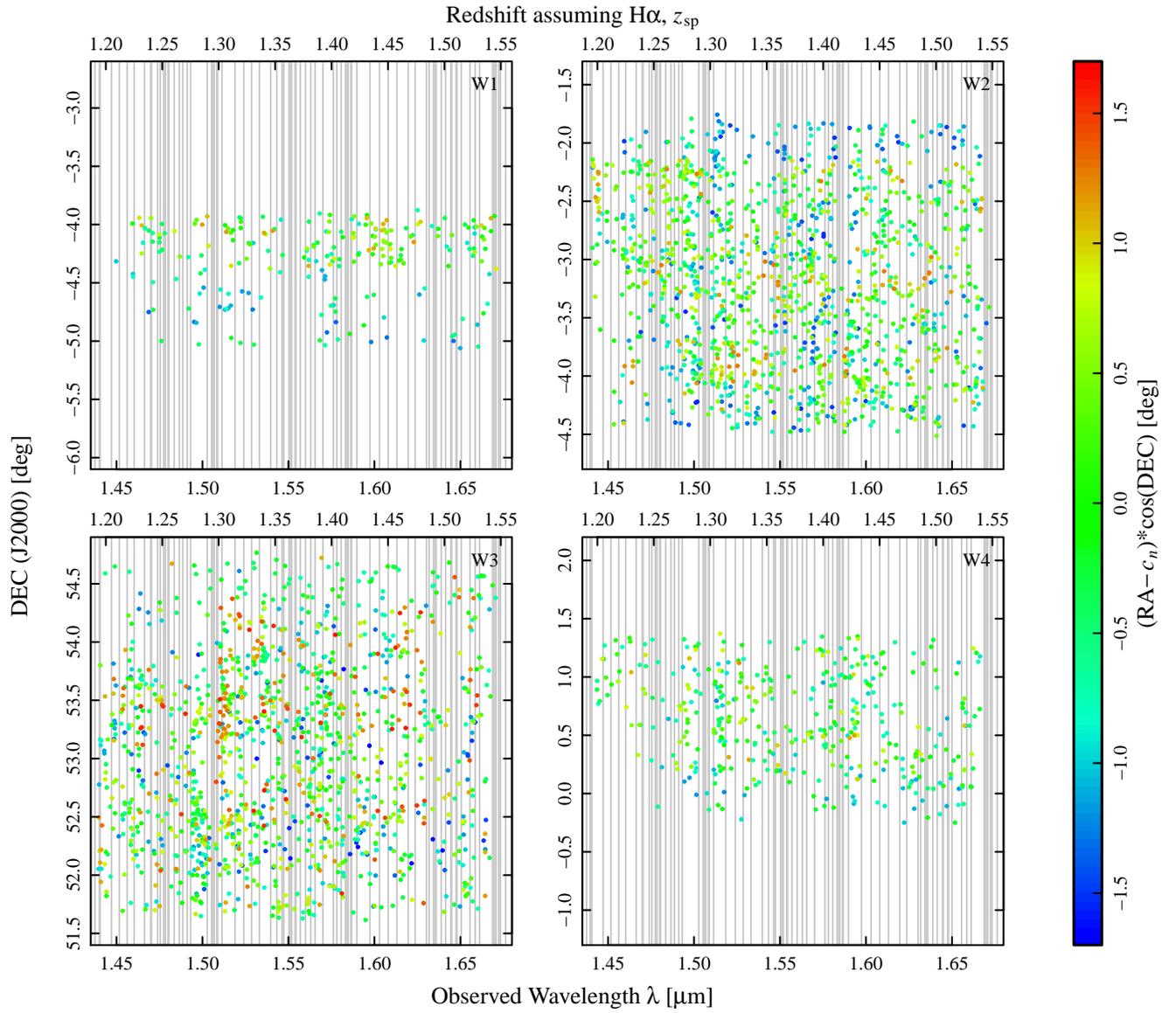}
\end{center}
   \caption{The same as Fig.~\ref{fig:RA_DEC} but for the galaxy
distribution in the DEC-wavelength space, with colors indicating
RA.  The RA zero points $c_i$ ($i$=1--4) are 35.5, 134.5, 214.5,
and 334.5 [deg] for the four fields of CFHTLS W1--4, respectively.
The wavelength ranges corresponding to the FMOS OH airglow masks
are indicated as the grey vertical stripes.}
\label{fig:z_DEC}
\end{figure*}

\begin{figure}[htp]
\begin{center}
  \includegraphics[width=0.4\paperwidth]{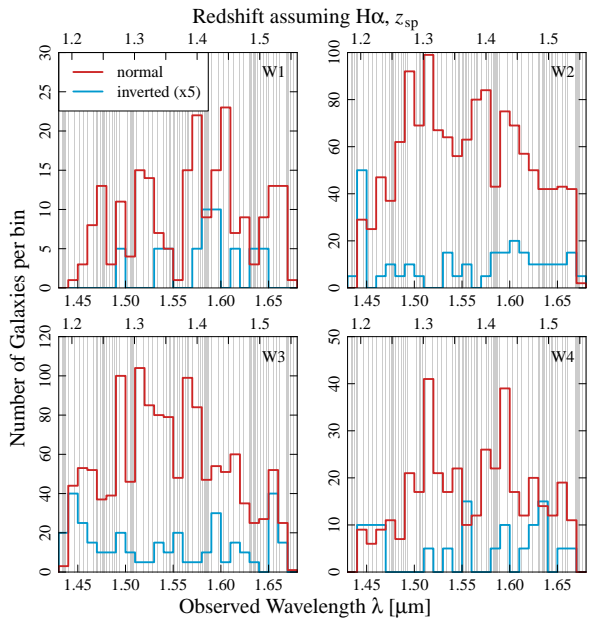}
\end{center}
  \caption{The wavelength distribution of the strongest line in a
    galaxy detected at $S/N \ge 4.5$ in the four CFHTLS-W fields.  The
    redshift assuming H$\alpha$ is also indicated at the top of the
    figure.  The red and blue histograms are for the line candidates
    in normal and inverted frames, respectively, and the latter is
    multiplied by a factor of five for the presentation purpose. Lines
    in inverted frames are all spurious lines, and they are used to
    estimate the spurious line fraction in normal frames.  The
    wavelength ranges corresponding to the FMOS OH airglow masks are
    indicated as the grey vertical stripes.  }
\label{fig:WL_distribution}
\end{figure}

\begin{figure*}[htp]
\begin{center}
  \includegraphics[width=0.82\paperwidth]{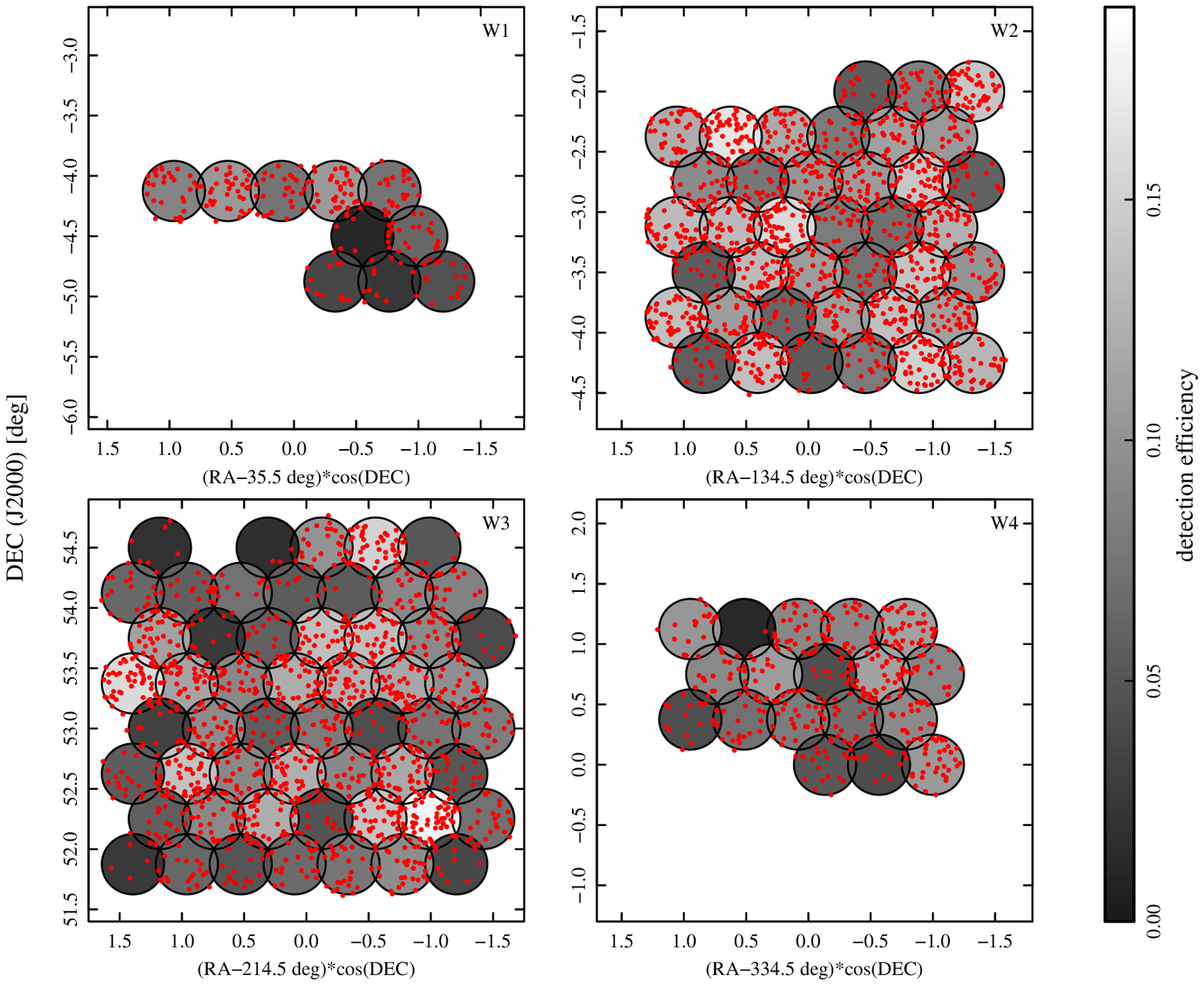}
\end{center}
  \caption{The detection efficiencies of emission line candidates,
    $N_{\rm elg}/N_{\rm obs}$ (i.e., the ratio of the number of
    galaxies with detected lines at $S/N \ge 4.5$ to that of all
    galaxies observed by FMOS), are shown for each FMOS FoV (circles)
    by greyscale.  Red dots represent the galaxies with detected
    emission lines. }
\label{fig:detection_efficiency}
\end{figure*}

\section{Basic Properties of Emission Line Galaxies}\label{Property}

Here we examine the basic properties of emission line galaxies
detected by the FastSound survey, using the sample of the highest
$S/N$ lines in a galaxy (i.e., $r_{\rm SN2}=1$) with the $S/N$
threshold of 4.5.  There are 3,299 galaxies, and for simplicity we
assume that all of these lines are H$\alpha$, though this sample
includes a small fraction of spurious noise detections ($\sim$5\%,
Paper I) and non-H$\alpha$ lines ($\sim$4\%, \S
\ref{section:contamination}).  This sample includes 11 emission lines
with large line widths of $\sigma_v > 500~\rm km \ s^{-1}$, whose host
galaxies are likely to be AGNs.  [Here the instrumental resolution is 
not deconvolved, which is typically $R \sim 2400$ in the H short+ 
wavelength range used in the FastSound observations (Paper I), 
corresponding to $\sigma_{v, \rm res} \sim 53$ km/s.] We excluded
these emission lines from our analysis, and our final sample comprises
3,288 galaxies.  These galaxies are distributed in the four CFHTLS-W
survey fields, with typical radial and tangential comoving lengths of
$\sim$700 and 100--200 Mpc, respectively.  The RA-Dec distributions of
these galaxies are shown in Fig.  \ref{fig:RA_DEC}, and the
Dec-wavelength distribution in Fig.  \ref{fig:z_DEC}, showing large
scale clustering features that will be discussed in our Paper IV
\citep{Okumura}.  The wavelength distributions are also shown as
histograms in Fig. \ref{fig:WL_distribution}.  The 3D large scale
clustering can be seen in these figures, though the line detection
efficiency is not uniform in different FoVs due to observing
conditions.  In Fig. \ref{fig:detection_efficiency}, we show the
RA-Dec map with the background color scale for the line detection
efficiency ($N_{\rm elg}/N_{\rm obs}$ in the FoV-information list).

\begin{figure}[htp]
\begin{center}
  \includegraphics[width=0.4\paperwidth]{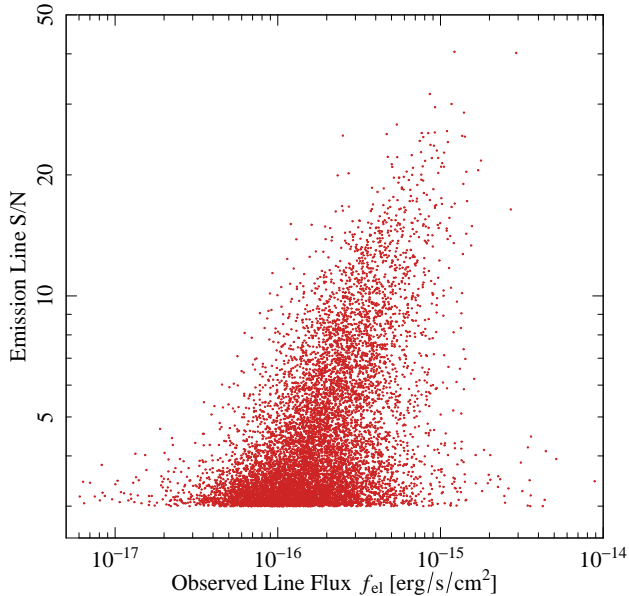}
\end{center}
  \caption{Correlation between the observed total emission line flux
    $f_{\rm el}$ of the strongest line in a galaxy ($S/N \ge 3$) and the
    line detection $S/N$.}
\label{fig:line_flux_SN}
\end{figure}

\begin{figure}[htp]
\begin{center}
\end{center}
  \includegraphics[width=0.4\paperwidth]{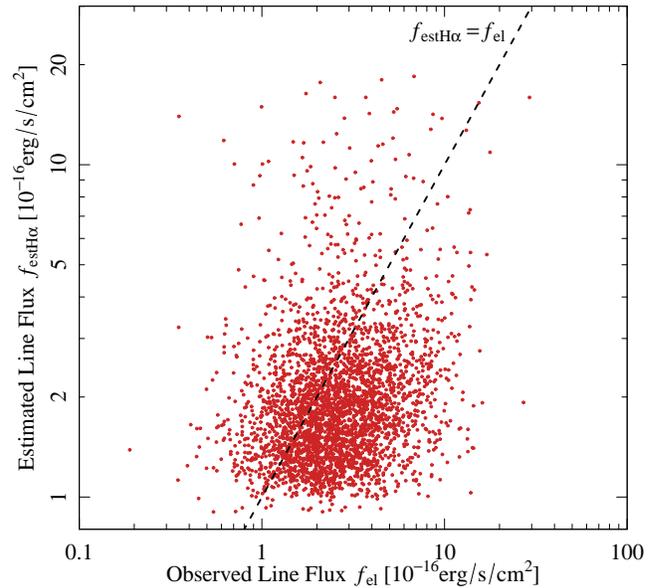}
  \caption{Correlation between the observed total emission line
flux $f_{\rm el}$ of the strongest line in a galaxy ($S/N \ge 4.5$)
and H$\alpha$ flux estimated by photometric SED fittings. 
}
\label{fig:flux_obs_est}
\end{figure}

\begin{figure}[htp]
\begin{center}
\end{center}
  \includegraphics[width=0.4\paperwidth]{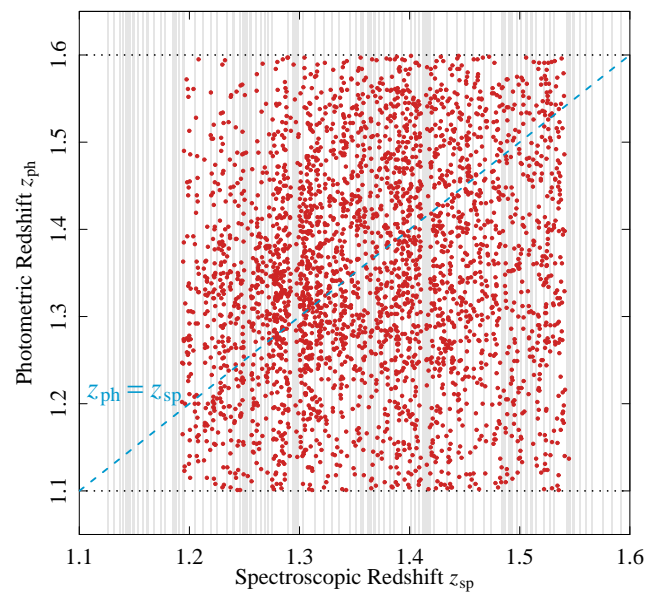}
  \caption{Photometric redshifts plotted against spectroscopic redshifts
    assuming that the strongest line in a galaxy is H$\alpha$. The
    FastSound target selection criteria is $1.1 < z_{\rm ph} <1.6$,
    which is indicated by dotted lines. The spectroscopic redshift
    ranges corresponding to the FMOS OH airglow masks are indicated as
    the grey vertical stripes. 
}
\label{fig:z_photoz}
\end{figure}

Fig. \ref{fig:line_flux_SN} shows the correlation between line $S/N$
and line flux $f_{\rm el}$. This plot indicates that the mean
sensitivity of FastSound is $\sim 1.6 \times 10^{-16} \rm \ erg
\ cm^{-2} s^{-1}$ at $S/N=5$, though scatter
around the mean relation is possible by varying observing conditions
and the effect of OH masks.  (Note that $f_{\rm el,raw}$ and $f_{\rm
  el}$ are integration over all wavelengths but $S/N$ is over regions
outside OH masks.)  We compare the observed flux $f_{\rm el}$ with the
estimated H$\alpha$ flux by photometric SED fittings for target
selection, in Fig. \ref{fig:flux_obs_est}.  Almost no correlation is
seen, most likely because of the uncertainty in the photometric SED
fitting using only 5 optical bands of CFHTLS-W.
Fig. \ref{fig:z_photoz} shows the correlation between photometric and
spectroscopic redshifts. The photometric redshifts are in the range of
$1.1 < z_{\rm ph} <1.6$ by the selection criteria, while spectroscopic
redshifts in $1.2 < z_{\rm sp} < 1.55$ limited by the observed
wavelength range. In this limited ranges, the scatter is large, again
because of the limited photo-$z$ accuracy (typical error of $\Delta
z_{\rm ph} \sim$0.15) based only on 5 optical bands.  The
uncertainties in the SED fittings are the main source of rather low
line detection efficiency ($\sim$ 10\%), but the adopted method is
still much better than no selection at all (see \cite{Tonegawaa}).

\begin{figure*}[htbp]
 \begin{minipage}{0.5\hsize}
  \begin{center}
   \includegraphics[width=0.4\paperwidth]{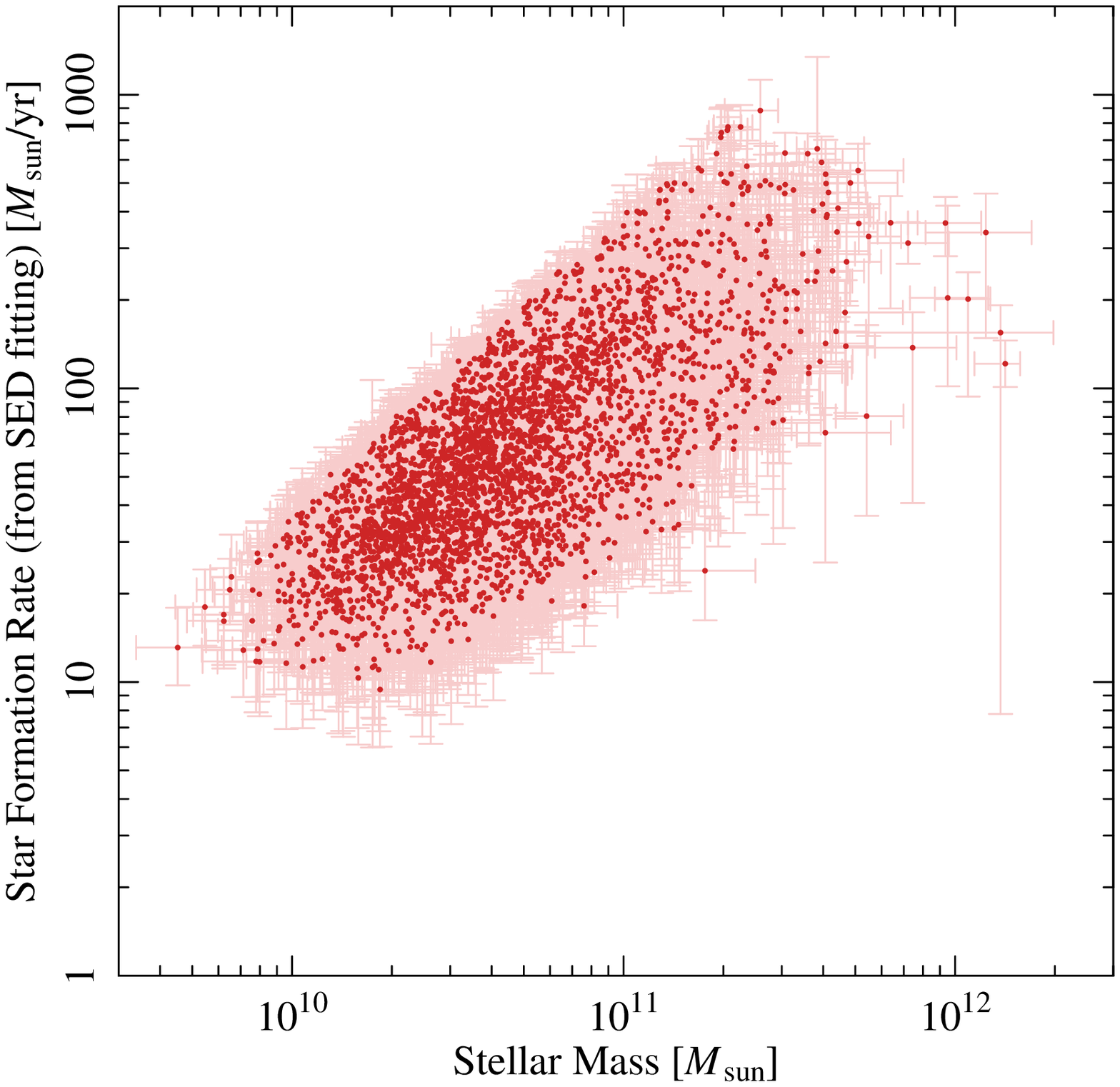}
  \end{center}
 \end{minipage}
 \begin{minipage}{0.5\hsize}
  \begin{center}
   \includegraphics[width=0.4\paperwidth]{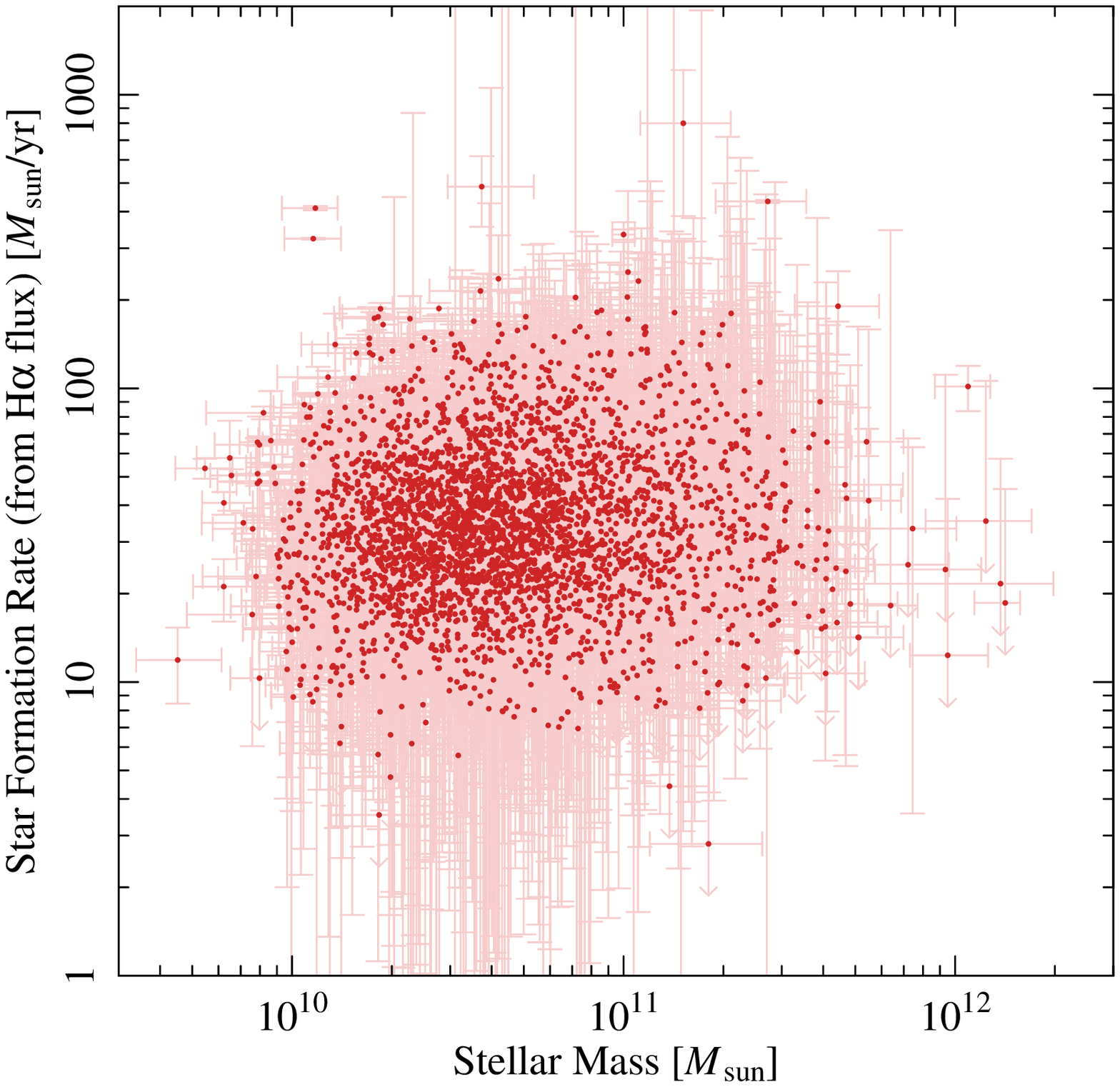}
  \end{center}
 \end{minipage}
   \hspace{3pt}
   \caption{Correlation between stellar mass ($M_*$) and SFR (left:
     estimated by SED fittings, right: by observed H$\alpha$ fluxes)
     for the host galaxies of emission lines detected by FastSound.
     In the left panel galaxies are distributed only in the region of
     $M_* \gtrsim $ SFR $\times$ 0.3 Gyr because of the SED model
     fitting method (see text). }
\label{fig:mass_SFR}
\end{figure*}

\begin{figure}[htp]
\begin{center}
  \includegraphics[width=0.4\paperwidth]{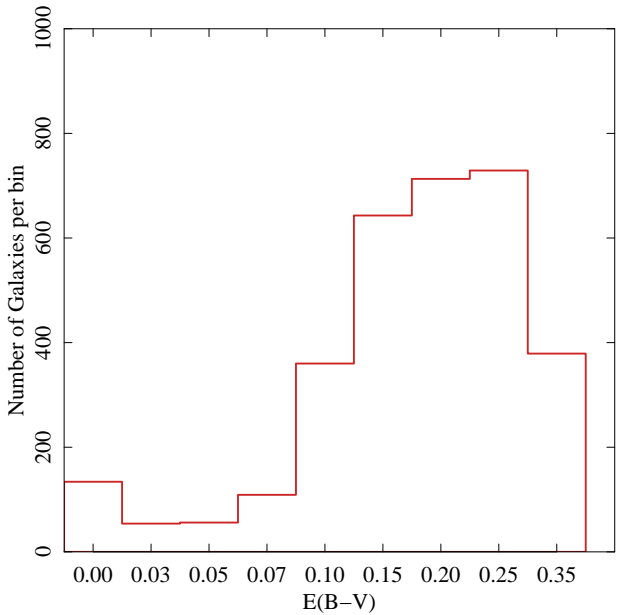}
\end{center}
  \caption{Histogram of $E(B-V)$ for the FastSound galaxies with
    detected emission lines.  Note that $E(B-V)$ intervals are not
    uniform, but the bins are set so that the bin centers correspond
    to the grid points in the model SED fitting (shown in the
    abscissa).  The range of $E(B-V)$ was limited to $E(B-V)<0.35$ in
    the SED fitting.  }
\label{fig:EBV}
\end{figure}

The correlation between SFR and stellar mass is shown in
Fig. \ref{fig:mass_SFR}, indicating that the typical ranges of these
quantities are 20--500 $M_\odot$/yr and $10^{10.0}$--$10^{11.3}
M_\odot$, respectively. Here, two panels are for two different SFR
estimates: one is by SED fittings (SFR$_{\rm SED}$), and the other is
by measured H$\alpha$ line fluxes and $E(B-V)$ from SED fittings
(SFR$_{\rm H\alpha}$).  The former was estimated by the photometric
SED fittings in the target selection processes (Paper I), using
templates of exponentially decaying SFR evolution and five optical
magnitudes in $u'g'r'i'z'$ bands.  Most galaxies were best-fitted with
approximately constant SFR up to the galaxy age $t_g$, i.e., the
exponential time scale $\tau$ much larger than $t_g$.  Since the SED
fittings are limited in the range of $t_g > t_{g,l} = 0.3$ Gyr, there
is a limit to the SFR$_{\rm SED}$-$M_*$ relation as $M_* > $ SFR$_{\rm
  SED}\, \times \, t_{g,l}$ , which can be seen in this figure. For
SFRs from observed H$\alpha$ fluxes, we followed the same relation
between SFR, H$\alpha$ emission line luminosity, and $E(B-V)$ as in
Paper I (eq. 5).  There is no clear correlation between $M_*$
and SFR$_{\rm H\alpha}$, but the mean values are broadly consistent
with the so-called main sequence of galaxies at similar redshifts
\citep{Karim,Speagle}.  The distribution of $E(B-V)$
estimated by the SED fittings for galaxies with detected lines is
shown in Fig. \ref{fig:EBV}.  Note that the fitting was limited in the
range of $E(B-V) < 0.35$.

\begin{figure}[htp]
\begin{center}
  \includegraphics[width=0.4\paperwidth]{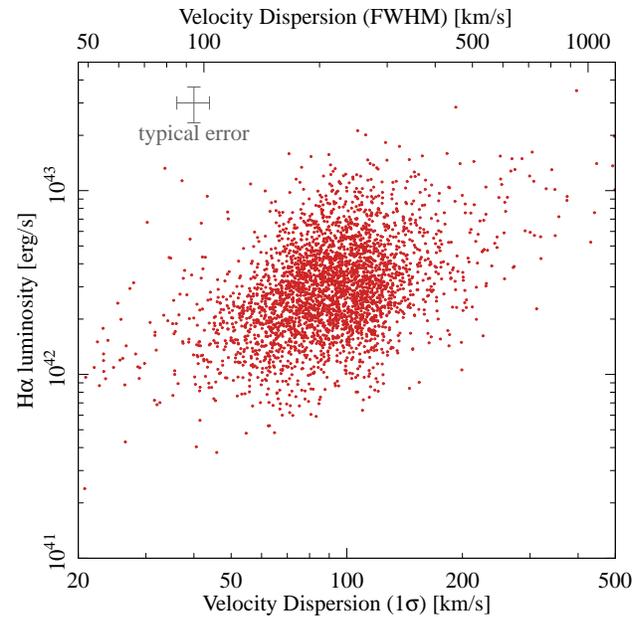}
\end{center}
  \caption{ Correlation between H$\alpha$ luminosity and velocity
    dispersion of FastSound galaxies. The velocity dispersions are not
    deconvolved with the instrumental spectral resolution ($\sigma \sim 50$
    km/s).}
\label{fig:LW_luminosity}
\end{figure}

We also plot observed H$\alpha$ luminosity versus velocity dispersion
for line-detected galaxies in Fig \ref{fig:LW_luminosity}.  H$\alpha$
luminosities of typical FastSound galaxies are in the range of
$10^{41.8}$--$10^{43.3}$ erg/s, which is the bright-end ($L_{\rm
  H\alpha} \gtrsim L_*$) of the H$\alpha$ luminosity function (LF) at
$z \sim 1.2$ \citep{Colbert}.  Here, velocity dispersion was
calculated from the observed line width $\sigma_\lambda$, and this is
not deconvolved with the instrumental resolution ($\sigma_{v, \rm res}
\sim 50$ km/s).  Indeed the majority of galaxies have $\sigma_v
\gtrsim \sigma_{v, \rm res}$.  A small fraction of galaxies with
$\sigma_v < \sigma_{v, \rm res}$ may be a result of low $S/N$ or
spurious noise detection.  We expect, on average, larger or more
massive systems would have larger velocity dispersions, as expected
from the general scaling relations for galaxies.  A positive
correlation is indeed seen in this figure, though a quantitative
analysis and interpretation are beyond the scope of this paper.

\begin{figure}[htp]
\begin{center}
  \includegraphics[width=0.4\paperwidth,angle=-90]{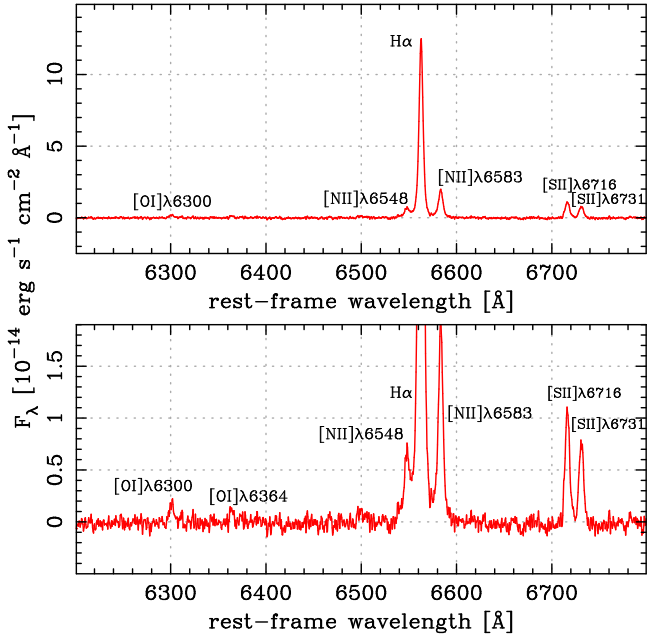}
\end{center}
  \caption{The stacked and continuum-subtracted spectrum of all 2719
    FastSound galaxies with detected emission lines ($S/N \ge 5$).
    The lower panel is the same as the top panel, but with different
    scale to clearly show the weak lines.  The excesses seen at 6375
    and 6502 {\AA} are likely due to H$\beta$ and [OIII]$\lambda$4959
when the [OIII]$\lambda$5007 line is confused to be H$\alpha$ (see text). 
}
 \label{fig:stacked}
\end{figure}

Finally, Fig. \ref{fig:stacked} shows the stacked and
continuum-subtracted spectrum of all the FastSound emission line
galaxies (2719 above $S/N = $ 5), assuming that the strongest line is
H$\alpha$.  
Along with the trivial H$\alpha$ line, the major lines
around H$\alpha$, i.e., [NII] and [SII] doublets, are clearly
detected. The even weaker [OI] doublet (about $\sim$1 \% level of
H$\alpha$ flux) is also seen. Table \ref{table:flux_stacked}
summarizes the line fluxes relative to H$\alpha$.

\begin{table}[htp]
\caption{The list of line flux relative to that of H$\alpha$ in the stacked spectrum shown in 
  Fig.~\ref{fig:stacked}. }
\label{table:flux_stacked}
\begin{center}
\begin{tabular}{cc}
\hline \hline
   Line &  relative flux \\
\hline
[OI]$\lambda$6300 & $0.015 \pm 0.003$\\

[OI]$\lambda$6364 & $0.008 \pm 0.002$\\

[NII]$\lambda$6548 & $0.081 \pm 0.004$\\

[NII]$\lambda$6583 & $0.177 \pm 0.003$\\

[SII]$\lambda$6717 & $0.095 \pm 0.003$\\

[SII]$\lambda$6731 & $0.069 \pm 0.003$\\

\hline \hline
\end{tabular}
\end{center}
\end{table}

\section{Contamination from Non-H$\alpha$ Emission Lines}
\label{section:contamination}

Though the majority of the emission lines detected in the FastSound
survey are expected to be H$\alpha$, it is crucial to estimate the
contamination fraction of other lines for precise scientific analyses.
For example the clustering signal in the RSD analysis is diluted by
large redshift errors, and because the errors are non-random a systematic
aliasing may be introduced.
Table \ref{table:line_list} shows 11 emission lines that are generally
found in star forming galaxies and hence can be confused with
H$\alpha$ lines, with a typical energy flux ratio against H$\alpha$.
Most FastSound galaxies were detected with a single emission line,
making it difficult to firmly identify redshifts. However, a fraction
of relatively bright FastSound galaxies were detected with multiple
lines, and we can estimate the contamination of non-H$\alpha$ lines by
the statistics of line identification for such galaxies. We report the
statistics of multiple line galaxies in FastSound in \S
\ref{section:multi-line-stat}, and in \S
\ref{section:multi-line-nHa-est} estimate non-H$\alpha$ fraction in
all the FastSound galaxies based on the results of \S
\ref{section:multi-line-stat}.  We will also make an independent
estimate of non-H$\alpha$ fraction from the stacked spectrum of the
FastSound galaxies in \S \ref{section:stacked-spec-nHa-est}.

\subsection{Emission Line Identification for Multiple-Line Objects}
\label{section:multi-line-stat}

Though the wavelength range of FastSound is rather narrow (1.44--1.67
$\mu$m), Table \ref{table:line_list} indicates that most lines can be
identified by nearby other lines if the sensitivity is good or
galaxies are bright, because most lines can be classified into a group
within a narrow wavelength range, i.e., H$\beta$-[OIII],
H$\alpha$-[NII]-[SII], and [SIII] systems.  The only exception is
[OII] at $z \sim$ 2.9--3.5, which may be detected as a single line
even if the galaxy is sufficiently bright, and hence the contamination
of [OII] cannot be examined by multiple-line galaxies.  (Actually
[OII] is a close doublet, but its separation, 220 km/s, is small
enough to be treated as a single line in the present catalog. See
\S\ref{section:emission-line-catalog}.)  However, the statistics of
the HiZELS narrow-band survey at a similar wavelength and sensitivity
implies that this contamination is expected to be negligible compared
with other contaminants such as [OIII] \citep{Sobral12}.  In fact,
using the [OII] LF measured at $z = 3.34$ \citep{Khostovan}, the
expected number of galaxies whose [OII] fluxes are detectable by the
FastSound sensitivity is 0.33\% of that of H$\alpha$ emitters at $z
\sim$ 1.2--1.5.  [Here, we used the extinction-uncorrected H$\alpha$
  LF parameters at $z = $ 0.9--1.5 reported by \citet{Colbert} for the
  number of H$\alpha$ emitters, which are in good agreement with those
  at $z = 1.47$ reported by \citet{Sobral13}.]  Therefore, even if the
target selection efficiency is the same for H$\alpha$ and [OII], the
contamination is only $\sim$0.3\%.  Furthermore, the actual
contamination should be lower than this, because our target selection
based on photometric redshifts should more efficiently remove [OII]
emitters at $z \sim 3$ than H$\alpha$ emitters.

Another source of contamination not shown in Table
\ref{table:line_list} is the Paschen series in near infrared bands,
among which the most significant for FastSound is expected to be
Pa$\beta$ (1.282 $\mu$m) at $z = $ 0.12--0.30. Assuming a flux ratio
of 0.1 for Pa$\beta$/H$\alpha$, which is valid for typical extinction
of $\sim$ 1 mag for H$\alpha$, we estimate the number of detectable
Pa$\beta$ emitters in this redshift range to be 4.3\% of H$\alpha$
emitters detectable by FastSound. Here, we used the same H$\alpha$ LF
parameters at $z=$1.2--1.5 as the previous paragraph, while the
extinction-uncorrected LF parameters for H$\alpha$ emitters at $z \sim
0.2$ are estimated by interpolating those for $z =$ 0.08 and 0.4
reported in \citet{Sobral13}. Furthermore, H$\alpha$ emitters should
be selected more efficiently as FastSound targets than Pa$\beta$
emitters, because of the selection by photo-$z$. According to the
photo-$z$ results of \citet{Tonegawaa} for galaxies with spec-$z$
using the same five optical bands of CFHTLS, the probability that
galaxies at $0 < z_{\rm sp} < 0.5$ are misidentified as galaxies at
$z_{\rm ph} > 1$ is less than 5\%.  Hence we conclude that the
contamination from Pa$\beta$ emitters is negligible.

We constructed the sample of multiple-line objects in FastSound by
requiring that a galaxy must have more than two lines detected at $S/N
\ge 3.0$, and at least one of them is detected at $S/N \ge 4.5$.
There are 1,105 galaxies satisfying this condition, which should be
compared with 3,288, the number of all FastSound galaxies with a line
detected at $S/N \ge 4.5$. We try to identify these multiple emission
lines by the wavelength ratio. The observed wavelength ratios
$\lambda_1 / \lambda_2$ ($\lambda_1 > \lambda_2$) are compared with
all the possible pairs of lines in Table \ref{table:line_list},
allowing an error of $\pm 5 \times 10^{-4}$, which is about four times
larger than the statistical 1$\sigma$ error expected from the typical
wavelength determination error of FastSound.  When more than three
emission lines are detected for a galaxy, we choose a pair of the two
brightest lines, and hence the number of emission line pairs
considered is the same as the number of multiple-line galaxies.

\begin{figure*}[htp]
\begin{center}
  \includegraphics[width=0.82\paperwidth]{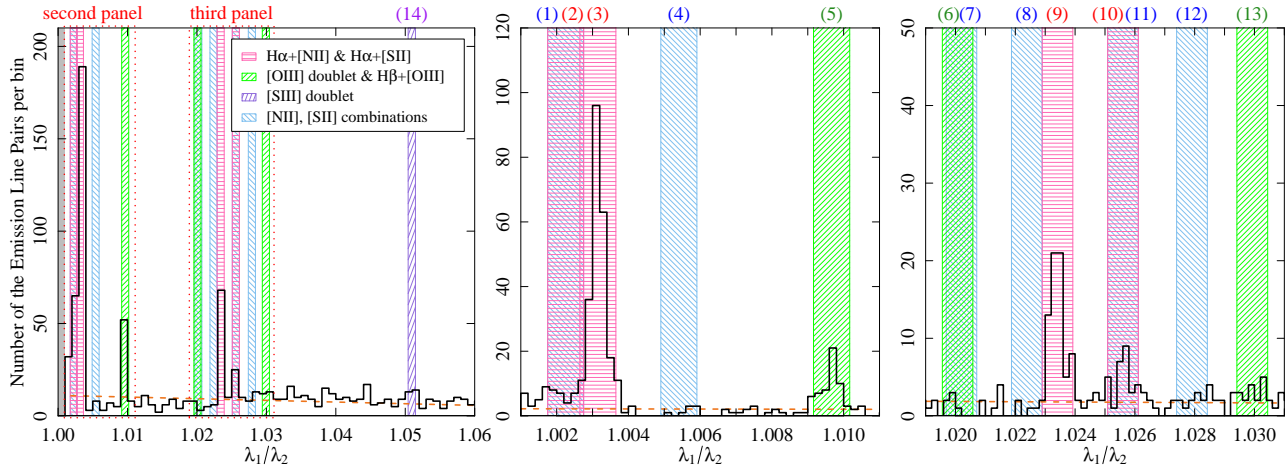}
\end{center}
  \caption{ The wavelength ratio distribution of the detected line
    pairs.  The left panel is showing the whole $\lambda_1/\lambda_2$
    range, and the middle and right panels are showing the close-up
    regions where many different species of line pairs are expected.
    The expected wavelength ratios for the possible line pairs listed
    in Table \ref{table:el_pair} are indicated by the vertical stripes
    [with the numbering (1)--(14) same as Table \ref{table:el_pair}].
    Pink, green, purple and blue stripes indicate H$\alpha$-X,
    H$\beta$-[OIII], [SIII] doublet, and [NII]-[SII] systems,
    respectively.  The dashed curve is the fit to the distribution
    after removing the regions of the expected line pairs, indicating
    the expected number of spurious line pair detections. 
}
 \label{fig:WL_ratio_distribution}
\end{figure*}

All possible 14 line pairs that can be detected in the FastSound
wavelength range (1.44--1.67 $\mu$m, i.e., $\lambda_1/\lambda_2 \leq
1.16$) are listed in Table \ref{table:el_pair}, and the numbers of
galaxies whose wavelength ratio is consistent with a line pair are
also shown. The histogram of $\lambda_1/\lambda_2$ is shown in
Fig. \ref{fig:WL_ratio_distribution}. A fraction of the detected line
pairs may be artefacts because spurious lines are included in the
FastSound line catalog, as discussed in Paper I.  This fraction can be
estimated in this histogram by the number of galaxies where no real
emission line pairs are expected. After removing the wavelength ratio
regions corresponding to the expected line pairs listed in Table
\ref{table:el_pair}, the $\lambda_1/\lambda_2$ histogram is fitted by
a linear function, and this is used to estimate the number of
artefacts. The real numbers of physical emission line pairs are
thus estimated by subtracting these numbers, as reported in Table
\ref{table:el_pair}.

A problem is that some line pairs have overlapping wavelength ratios
with the above allowance interval of $5 \times 10^{-4}$. There are
three such cases: (A) H$\alpha$/[NII]$\lambda$6548 = 1.00225 and [SII]
doublet 6731/6717 = 1.00214, (B) [OIII]$\lambda$4959/H$\beta$ =
1.02007 and [SII]$\lambda$6717/[NII]$\lambda$6583 = 1.02020, and (C)
[SII]$\lambda$6717/[NII]$\lambda$6548 = 1.02560 and
[SII]$\lambda$6731/H$\alpha$ = 1.02572.  For the cases (A) and (C), it
is expected that the majority of the detected galaxies should be
H$\alpha$+[NII] and H$\alpha$+[SII] for (A) and (C), respectively,
because H$\alpha$ is the strongest line. Furthermore, if the [SII]
doublet and [NII]$\lambda$6548+[SII]$\lambda$6717 are significantly
detected in the line pair sample considered here, we expect that the
[NII] doublet and [NII]$\lambda$6548+[SII]$\lambda$6731 should also be
detected because of similar line fluxes. However these pairs are not
significantly detected in the line pair sample (see Table
\ref{table:line_list}).  Therefore, we regard all the emission line
pairs detected in the cases (A) and (C) as H$\alpha$ plus [NII] or
[SII].  For the case (B), no significant excess was found at this
wavelength ratio, and hence we can safely ignore this.

Then the significantly detected line pairs in Table
\ref{table:el_pair} can be classified into the two categories: one is
pairs including H$\alpha$ (denoted as H$\alpha$-X hereafter) and the
other is the [OIII] doublets. Evidence for the
H$\beta$-[OIII]$\lambda$5007 and [SIII] doublet pairs is found,
though statistical significance is about 2 $\sigma$ level (chance
probability of $\sim$5\%).  The number of galaxies whose line pairs
are identified as H$\alpha$-X is 312.1 when the expected number of
spurious detections is subtracted. Note that this number is slightly
different from the simple sum of the corrected number $n-n_n$ in Table
\ref{table:el_pair} because the numbers in this Table are multiply
counted when the wavelength ratio ranges are overlapping.  This number
should be compared with the detected non-H$\alpha$ pairs: 39.7, 6.4,
and 5.2 for the [OIII] doublet, H$\beta$-[OIII]$\lambda$5007, and
[SIII] doublet, respectively. 

\begin{figure*}[htp]
\begin{center}
  \includegraphics[width=0.82\paperwidth]{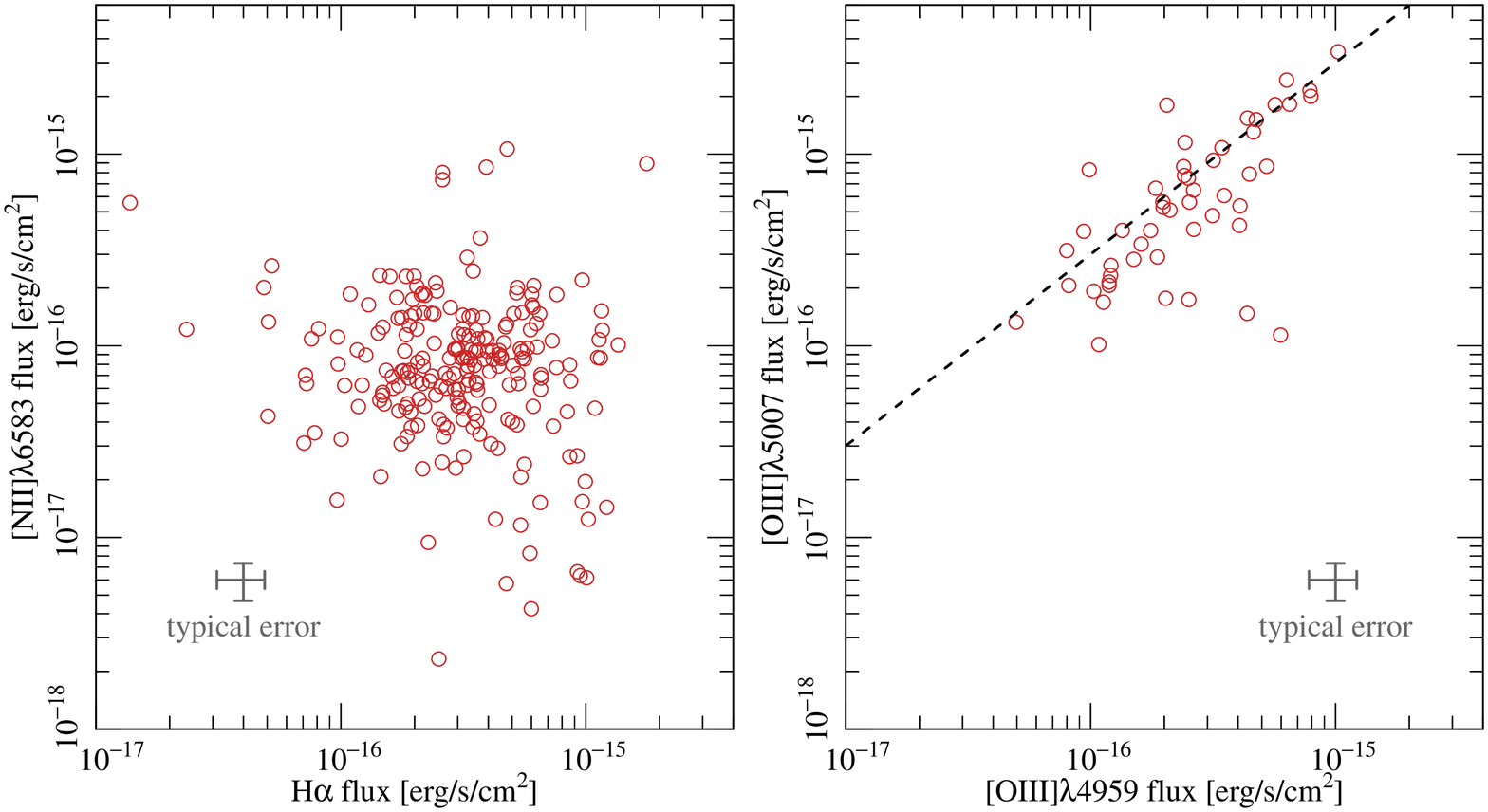}
\end{center}
  \caption{ Line flux correlations for the identified
    H$\alpha$+[NII]$\lambda$6583 pairs (left) and [OIII] doublets
    (right).  The dashed line shows the ratio (3:1) expected from
atomic physics for [OIII] doublet.
Each panel should include about 10 fake pairs caused
    by noise (see Table~\ref{table:el_pair}), which explain
galaxies with anomalous line ratios. 
}
 \label{fig:2line_flux}
\end{figure*}

Fig. \ref{fig:2line_flux} shows the correlation plots of the line pair
fluxes for H$\alpha$-[NII]$\lambda$6583 and the [OIII] doublets.  Only
weak correlation can be seen for H$\alpha$-[NII]$\lambda$6583, which
is not unreasonable because the line flux ratio depends on metallicity
and the ionization status of a star forming galaxy. On the other hand,
we expect a tighter correlation for the [OIII] doublets around the
expected value of 3:1 from atomic physics, and this is indeed observed
in this figure.

%
\begin{table}[htp]
\small
\caption{The list of major emission lines expected for star forming
  galaxies.  The redshift range corresponding to the FastSound
  wavelength range (1.44--1.67~$\mu$m) and typical line strength
  (energy flux normalized by H$\alpha$) are also shown.
}
\label{table:line_list}
\begin{center}
\begin{tabular}{lccc}
\hline \hline
   Emission line &  wavelength~[$\mu$m]\footnotemark[$*$] & $z$ & relative flux\\ 
\hline
[OII]$\lambda$3727 &  $0.3727$ &   $2.86$--$3.48$ & $0.47$\footnotemark[$\dagger$] \\ 

H$\beta$ &  $0.4863$ &   $1.96$--$2.43$ & $0.15$\footnotemark[$\dagger$] \\ 

[OIII]$\lambda$4959 &  $0.4960$ &   $1.90$--$2.37$ & $0.07$\footnotemark[$\dagger$] \\ 

[OIII]$\lambda$5007 &  $0.5008$ &   $1.88$--$2.33$ & $0.27$\footnotemark[$\dagger$] \\ 

[NII]$\lambda$6548 &  $0.6550$ &   $1.20$--$1.55$ & $0.17$\footnotemark[$\dagger$] \\ 

H$\alpha$ &  $0.6565$ &   $1.19$--$1.54$ & $1$\ \ \ \ \ \ \ \ \\ 

[NII]$\lambda$6583 &  $0.6585$ &   $1.19$--$1.54$ & $0.46$\footnotemark[$\dagger$] \\ 

[SII]$\lambda$6717 &  $0.6718$ &   $1.14$--$1.49$ & $0.19$\footnotemark[$\dagger$] \\ 

[SII]$\lambda$6731 &  $0.6733$ &   $1.14$--$1.48$ & $0.14$\footnotemark[$\dagger$] \\ 

[SIII]$\lambda$9069 &  $0.9071$ &   $0.59$--$0.84$ & $0.1$\footnotemark[$\ddagger$]\ \ \ \\ 

[SIII]$\lambda$9531 &  $0.9533$ &   $0.51$--$0.75$ & $0.2$\footnotemark[$\ddagger$]\ \ \ \\ 

\hline\hline
\end{tabular}
\end{center}
{\footnotesize
 \footnotemark[$*$] Rest-frame wavelength measured in a vacuum \\
 \footnotemark[$\dagger$] \citet{Glazebrook03} \\
 \footnotemark[$\ddagger$] \citet{Garnett} 
 }
\end{table}

\begin{table*}[htp]
\small
\caption{
The result of the line identification for the emission line
pairs found in the 1,105 FastSound galaxies. 
} 
\begin{center}
\begin{tabular}{rlllrrrl}
\hline \hline
  & line (shorter) & line (longer) & \footnotesize{wavelength ratio} & number $n$ & noise\footnotemark[$*$] $n_\mathrm{n}$ & \footnotesize{number (corrected)}\footnotemark[$\dagger$]   & \footnotesize{chance probability}\footnotemark[$\ddagger$] \\
\hline
(1) & [SII]$\lambda$6717 & [SII]$\lambda$6731 & $1.00214$\footnotemark[$\S$] & $35$ & $10.9$ & $24.1^{+7.0}_{-5.9}$ & $< 10^{-5}$ \\ 
(2) & [NII]$\lambda$6548 & H$\alpha$ & $1.00225$\footnotemark[$\S$] & $37$ & $10.9$ & $26.1^{+7.1}_{-6.1}$ & $< 10^{-5}$ \\ 
(3) & H$\alpha$ & [NII]$\lambda$6583 & $1.00315$ & $226$ & $10.9$ & $215.1^{+16.1}_{-15.0}$ & $< 10^{-5}$ \\ 
(4) & [NII]$\lambda$6548 & [NII]$\lambda$6583 & $1.00541$ & $6$ & $10.7$ & $< 3.78$ & $0.95$ \\ 
(5) & [OIII]$\lambda$4959 & [OIII]$\lambda$5007 & $1.00966$ & $50$ & $10.3$ & $39.7^{+8.1}_{-7.0}$ & $< 10^{-5}$ \\ 
(6) & H$\beta$ & [OIII]$\lambda$4959 & $1.02007$\footnotemark[$\|$]  & $6$ & $9.5$ & $< 3.78$ & $0.91$ \\ 
(7) & [NII]$\lambda$6583 & [SII]$\lambda$6717 & $1.02020$\footnotemark[$\|$]  & $5$ & $9.5$ & $< 3.78$ & $0.96$ \\ 
(8) & [NII]$\lambda$6583 & [SII]$\lambda$6731 & $1.02238$ & $5$ & $9.3$ & $< 3.78$ & $0.95$ \\ 
(9) & H$\alpha$ & [SII]$\lambda$6717 & $1.02341$ & $67$ & $9.2$ & $57.8^{+9.2}_{-8.2}$ & $< 10^{-5}$ \\ 
(10) & H$\alpha$ & [SII]$\lambda$6731 & $1.02560$\footnotemark[$\#$]  & $26$ & $9.0$ & $17.0^{+6.2}_{-5.1}$ & $< 10^{-5}$ \\ 
(11) & [NII]$\lambda$6548 & [SII]$\lambda$6717 & $1.02572$\footnotemark[$\#$]  & $24$ & $9.0$ & $15.0^{+6.0}_{-4.9}$ & $0.00002$ \\ 
(12) & [NII]$\lambda$6548 & [SII]$\lambda$6731 & $1.02791$ & $11$ & $8.8$ & $2.2^{+4.4}_{-2.2}$ & $0.27$ \\ 
(13) & H$\beta$ & [OIII]$\lambda$5007 & $1.02993$ & $15$ & $8.6$ & $6.4^{+5.0}_{-3.8}$ & $0.031$ \\ 
(14) & [SIII]$\lambda$9069 & [SIII]$\lambda$9531 & $1.05094$ & $12$ & $6.8$ & $5.2^{+4.6}_{-3.4}$ & $0.046$ \\ 
\hline \hline
\end{tabular}
\label{table:el_pair}
\end{center}
{\footnotesize
      \footnotemark[$*$] 
The expected number of spurious pairs originating from noise. \\
      \footnotemark[$\dagger$]
The number corrected for the spurious pair detection rate,
i.e., $n - n_{\rm n}$, with 1$\sigma$ statistical errors. 
The upper bound is given at 2$\sigma$, assuming the
observed pairs are all spurious and hence no physical
line pairs were detected. \\
      \footnotemark[$\ddagger$] 
The probability of finding the observed number or more of
line pairs by noise events under the Poisson statistics. \\
      \footnotemark[$\S$,$\|$,$\#$] These pairs are indistinguishable 
  due to the very close values of wavelength ratio.}
\end{table*}
%
%
%

\subsection{Contamination Rate Estimation for Single Line Objects}
\label{section:multi-line-nHa-est}

\begin{figure}[htp]
\begin{center}
  \includegraphics[width=0.4\paperwidth]{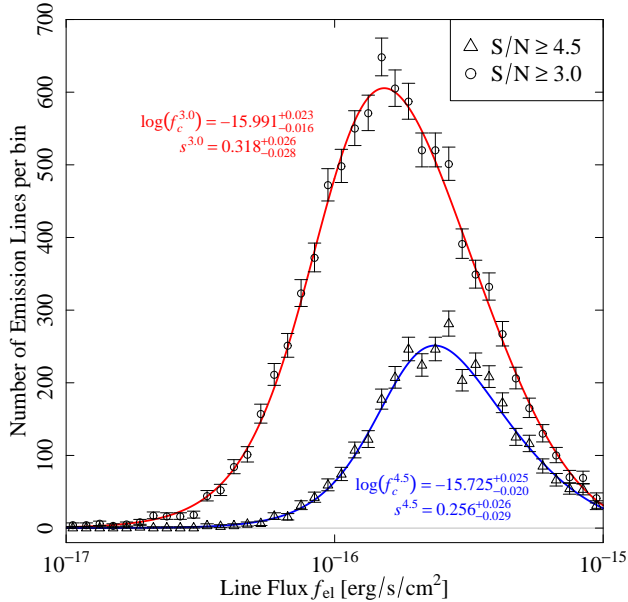}
\end{center}
  \caption{ Line flux distributions of the emission lines with the
    $S/N$ thresholds of 4.5 (triangles) and 3.0 (circles). The
    best-fit functions are shown with blue and red curves.  
}
 \label{fig:flux_fit}
\end{figure}

Based on the statistics of multiple emission lines reported in the
previous section, we estimate the contamination of non-H$\alpha$ lines
in the whole FastSound emission line catalog, in which most galaxies
have only one line. The only line pair that was significantly detected
except for those related to the H$\alpha$-[NII]-[SII] system is the
[OIII] doublet. First we model the line detection completeness (i.e.,
the probability for an emission line to be detected as a single line),
as a function of the total emission line flux $f_{\rm el}$, 
\begin{equation}
C_x(f_{\rm el}) = \frac{1} {1+(f_{\rm el}/f_c)^{-1/s}}, \label{eq:cutoff}
\end{equation}
where $x$ is the line detection threshold $S/N$, and $f_c$ and $s$ are
the fitting parameters.  For the line detection thresholds $S/N$ = 3.0
and 4.5, we estimated $f_c$ and $s$ by fitting the expected line flux
distribution, $C(f_{\rm el}) n(f_{\rm el})$ to the observed $f_{\rm
  el}$ distribution of the whole FastSound line catalog. Here,
$n(f_{\rm el})$ is the true differential flux counts per unit $f_{\rm
  el}$, and we assumed the form of $n(f_{\rm el})$ to be the Schechter
form, i.e., $n(f_{\rm el}) \propto (f_{\rm el}/f_*)^\alpha
\exp(-f_{\rm el}/f_*)$ with $\alpha$ and $f_*$ as free parameters. As
shown in Fig.~\ref{fig:flux_fit}, we got sufficiently good fits, and
obtained $(\log_{10}f_c, s)$ = ($-15.991^{+0.023}_{-0.016}$,
$0.318^{+0.026}_{-0.028}$) and ($-15.725^{+0.025}_{-0.020}$,
$0.256^{+0.026}_{-0.029}$) for $C_{3.0}$ and $C_{4.5}$, respectively,
where $f_c$ is in $\rm erg \ cm^{-2}\ s^{-1}$. Here, 
$\alpha$ and $f_*$ have been marginalized to estimate the
errors of $f_c$ and $s$. The best-fit values for these
are $(\log_{10} f_*, \alpha) = (-15.334, -1.637)$
and $(-15.535, -1.164)$, respectively. 

Then we can calculate the expected number of [OIII] doublets
with the line detection thresholds of $S/N = 4.5$ and 3.0
for the stronger and weaker lines, as adopted in the analysis
of the previous section, by
\begin{equation}
N_{\rm [OIII], double} = 
\int C_{3.0}(f_{\rm el}/r) \, C_{4.5}(f_{\rm el}) \, n_{\rm [OIII]}(f_{\rm el}) \, df_{\rm el} \ ,
\end{equation}
where $r$ is the line flux ratio of the line pair ($r=3$ for [OIII]
doublets), and $n_{\rm [OIII]}(f_{\rm el})$ is the flux counts for the stronger
line of the [OIII] doublets.  The expected number of all [OIII] lines
including those detected as a single line in the whole FastSound
sample is
\begin{equation}
N_{\rm [OIII], single} = \int C_{4.5}(f_{\rm el}) \, n_{\rm [OIII]}(f_{\rm el}) \, df_{\rm el}
\end{equation}
Therefore, if we know the form of $n_{\rm [OIII]}(f_{\rm el})$, we can
calculate $R \equiv N_{\rm [OIII], double} / N_{\rm [OIII], single}$,
and from the observed number of $N_{\rm [OIII], double}$, we can
estimate $N_{\rm [OIII], single}$.

We cannot simply use the observed flux counts of the whole FastSound
catalog for $n_{\rm [OIII]}$, because most of the lines are
H$\alpha$. Instead, we calculate $n_{\rm [OIII]}$ assuming the [OIII]
LF at $z =$1.9--2.3 to be the Schechter form with $L_* =
10^{42.91}\ {\rm erg\ s}^{-1}$ and $\alpha = -1.67$, which are
inferred from the observational estimates at $z =$ 1.5--2.3 by
\citet{Colbert}.  Here, a factor of 3/4 is multiplied to $L^*$ of
\citet{Colbert} to correct the unresolved doublet luminosity into that
of the stronger one. Then we found $R = 0.28^{+0.06}_{-0.06}$ (error
coming from those of $f_c$ and $s$), and hence our estimate of [OIII]
lines in the whole FastSound line catalog ($S/N \ge 4.5$) is
$4.35^{+1.51}_{-1.13}$\%, where the error includes those of $R$ and
Poisson statistics about the number of galaxies detected by [OIII]
doublets, 39.7$^{+8.1}_{-7.0}$ (Table \ref{table:el_pair}).
\citet{Colbert} also presented another LF parameter fit where $\alpha$
is fixed to $-1.5$, motivated by the $\alpha$ value inferred from
lower redshift data.  We found that the [OIII] contamination fraction
changes only by a factor of 1.039 when this LF parameter set is
assumed. If we adopt $\alpha =-1$ and $\alpha = -2$, the fraction
becomes $3.30^{+0.92}_{-0.74}$ and $5.08^{+2.00}_{-1.43}$\%, respectively.

\subsection{Estimation from the Stacked Spectrum}
\label{section:stacked-spec-nHa-est}

We can use the stacked spectrum in Fig. \ref{fig:stacked} to estimate
the contamination of the [OIII] line.  If the stronger line of the
[OIII] doublets ([OIII]$\lambda$5007) is confused as a H$\alpha$ line,
we expect the other line ([OIII]$\lambda$4959) appearing at $6565
\times 4960/5008 \sim 6502$ {\AA}, where wavelengths are those in
vacuum (see also Table \ref{table:line_list}).  
The stacked spectrum
indeed shows an excess at this wavelength.  The measured line ratio of
the 6502 {\AA} excess to H$\alpha$ is $(9.5 \pm 3.9) \times
10^{-3}$. Since [OIII]$\lambda$5007 is stronger than $\lambda$4959 by
an exact factor of 3, we get an estimate of 2.9$\pm$1.2\% as the
contamination fraction of [OIII] doublets in the total H$\alpha$ flux
of the stacked spectrum. Neglecting the difference of the flux
distribution, a similar fraction is expected for the number of
galaxies detected by [OIII] rather than H$\alpha$.  This result is in
nice agreement with that derived from the multiple-line statistics and
luminosity functions in the previous section.

It should be noted that the stacked spectrum also shows a marginal
excess at 6375~{\AA}, which may be a result of H$\beta$ when
[OIII]$\lambda$5007 is confused to be H$\alpha$. However, this excess
is not as large as that at 6502 {\AA}, indicating that the mean
H$\beta$ flux is fainter than [OIII]$\lambda$4959.

\section{Conclusion}\label{Conclusion}

This is the second paper of the series of the FastSound project, a
cosmological redshift survey that aims to detect redshift space
distortion (RSD) in clustering of galaxies at $z\sim 1.4$. The
basic concept and survey design were described in Paper I
\citep{Tonegawac}.

In \S \ref{ELcatalog}, we presented in detail the FastSound catalog of
emission line candidates and their host galaxies detected by the
FastSound survey.  The FastSound data set consists of the three
tables: {\it the FoV-information list} for the observational records
and galaxy/line statistics in each FMOS FoV, {\it the galaxy catalog}
for information of all galaxies selected as FastSound targets, and
{\it the emission line catalog} for all line candidates detected by
the survey.  The FastSound catalog contains $\sim$3,300 galaxies with
at least one emission line detected at $S/N \ge 4.5$, corresponding to
the total (fiber-aperture corrected) line flux limit of about $1.6
\times 10^{-16}~\rm erg \ cm^{-2} s^{-1}$.  The catalog is
already open to the public.

Because of the strength of H$\alpha$ lines and our target selection
for galaxies expected to have strong H$\alpha$ flux based on
photometric SED fittings, more than 90\% of the detected lines are
expected to be H$\alpha$ \citep{Tonegawaa}.  We presented basic
physical properties of the lines and host galaxies in \S
\ref{Property}, assuming that the strongest line in a galaxy is always
H$\alpha$.  The 3D distributions of the FastSound galaxies in the four
CFHTLS Wide fields are visualized, clearly showing large scale
clustering in the four boxes whose radial and tangential comoving
length are typically 700 and 200 Mpc, respectively.  Typical FastSound
H$\alpha$ emitters have H$\alpha$ luminosities of
$\sim10^{42}$--$10^{43}$~erg/s, SFRs of $\sim$20--500 $M_\odot$/yr,
and stellar masses of $\sim 10^{10.0}$--$10^{11.3}$ $M_\odot$.

Though the majority of FastSound emission lines are expected to be
H$\alpha$, a quantitative estimate of non-H$\alpha$ contamination is
crucial for the primary scientific purpose of FastSound, because
non-H$\alpha$ contamination would result in damping of the clustering
signal. Therefore we examined the galaxies with multiple line
candidates, and from the wavelength ratio of line pairs, we identified
the lines of $\sim$350 FastSound galaxies. It was found that about
88\% of these are the H$\alpha$-[NII]-[SII] system, and the majority
of the remaining 12\% are the [OIII]$\lambda \lambda$4959,5007 doublet
at $z \sim$ 1.9--2.3.  Galaxies with H$\beta$-[OIII]$\lambda$5007 and
[SIII]$\lambda\lambda$9069,9531 pairs are also detected with smaller
statistical significance, but they are negligible compared with those
identified by the [OIII] doublet.  No other line pairs were clearly
detected. From these statistics, and combined with the [OIII]
LF at $z \sim 2$, we estimated the contamination of
[OIII] doublets in the full FastSound catalog of galaxies with at
least one detected emission line to be $4.35^{+1.51}_{-1.13}$\%. 
As an
independent estimate, we calculated this contamination rate from the
analysis of the stacked spectrum of 2719 FastSound galaxies, which
resulted in a consistent value of 2.9$\pm$1.2\%.  The contamination by
[OII]$\lambda$3727 line emitters cannot be examined only by the
FastSound data set, but we showed that it is also negligible based on
the recent studies of [OII] LF at high redshifts.

The forthcoming papers will discuss the metallicity evolution at $z
\sim 1.4$ (Paper III, \cite{Yabe15}), the RSD measurement and a new
constraint on the structure growth rate (Paper IV, \cite{Okumura}), 
and various other topics. 

\bigskip

The FastSound project was supported in part by MEXT/JSPS KAKENHI Grant
Numbers 19740099, 19035005, 20040005, 22012005, and 23684007.
KG acknowledges support from ARC Linkage International Fellowship
LX0989763. 
AB gratefully acknowledges the hospitality of the Research School of Astronomy \&
Astrophysics at the  Australian National University, Mount Stromlo, Canberra 
where some of this work was done under the Distinguished Visitor scheme.

\end{document}